\documentclass[preprint2,11pt]{aastex}

\slugcomment{Accepted to AJ for publication May 2004}

\shorttitle{Seven New 2MASS T Dwarfs}
\shortauthors{Burgasser et al.}

\begin{document}

\title{The 2MASS Wide-Field T Dwarf Search. III. Seven New T Dwarfs and Other Cool Dwarf Discoveries}

\author{
Adam J.\ Burgasser\altaffilmark{1,2,3}, Michael W.\ McElwain\altaffilmark{1,3},
J.\ Davy Kirkpatrick\altaffilmark{4}, Kelle L.\ Cruz\altaffilmark{3,5},
Chris G.\ Tinney\altaffilmark{6}, \& I.\ Neill Reid\altaffilmark{3,7}
}

\altaffiltext{1}{Department of Physics \& Astronomy,
University of California
at Los Angeles, Los Angeles,
CA, 90095-1562; adam@astro.ucla.edu, mcelwain@astro.ucla.edu}
\altaffiltext{2}{Hubble Fellow}
\altaffiltext{3}{Visiting Astronomer at the Infrared Telescope Facility, which is operated by
the University of Hawaii under Cooperative Agreement NCC 5-538 with the National Aeronautics
and Space Administration, Office of Space Science, Planetary Astronomy Program.}
\altaffiltext{4}{Infrared Processing and Analysis Center, M/S 100-22,
California Institute of Technology, Pasadena, CA 91125; davy@ipac.caltech.edu}
\altaffiltext{5}{Department of Physics \& Astronomy, University of Pennsylvania, 209 South 33rd Street,
Philadelphia, PA 19104; kelle@sas.upenn.edu}
\altaffiltext{6}{Anglo-Australian Observatory, P.O. Box 296. Epping, NSW 1710, Australia;
cgt@aaoepp.aao.gov.au}
\altaffiltext{7}{Space Telescope Science Institute, 3700 San
Martin Drive, Baltimore, MD 21218; inr@stsci.edu}

\begin{abstract}
We present the discovery of seven new T dwarfs identified in
the Two Micron All Sky Survey. Low-resolution (R$\sim$150)
0.8--2.5 $\micron$ spectroscopy obtained with the IRTF SpeX
instrument reveal the characteristic H$_2$O and
CH$_4$ bands in the spectra of these brown dwarfs. Comparison to
spectral standards observed with the same instrument enable us to
derive classifications of T3 to T7 for the objects in this sample.  Moderate-resolution
(R$\sim$1200) near-infrared spectroscopy for a subset of these discoveries
reveal \ion{K}{1} line strengths consistent with previously observed trends with spectral type.
Follow-up imaging observations provide proper motion measurements
for these sources, ranging from $<$ 0$\farcs$1 to 1$\farcs$55 yr$^{-1}$.
One object, 2MASS 0034+0523, has a
spectrophotometric distance placing it within 10 pc of
the Sun.  This source also exhibits a depressed K-band peak reminiscent
of the peculiar T dwarf 2MASS 0937+2931, and may be a metal-poor or old, high-mass
brown dwarf.  We also present
low resolution SpeX data for a set of M and L-type dwarf, subdwarf, and giant comparison
stars used to classify 59 additional candidates identified as background stars.
These are primarily M5-M8.5 dwarfs, many exhibiting \ion{H}{1} Paschen $\gamma$, but
include three candidate ultracool M subdwarfs
and one possible early-type L subdwarf.
\end{abstract}


\section{Introduction}

T dwarfs are a spectral class of brown dwarfs distinguished by the
presence of CH$_4$, H$_2$O, and H$_2$ collision-induced
absorption (CIA) in the near-infrared
\citep{kir99,me02a,geb02}, and heavily pressure-broadened \ion{K}{1} and \ion{Na}{1} absorption
at optical wavelengths \citep{tsu99,bur00,me03d}.  These objects have effective
temperatures (T$_{eff}$s) ranging from $\sim$1300 K
at the transition between L and T dwarfs \citep{kir00,stp01,dah02,vrb04}
to $\sim$750 K for the latest-type T dwarf 2MASS 0415$-$0935 \citep{me02a,vrb04}.
T dwarfs therefore comprise the coldest and intrinsically faintest brown dwarfs
currently known, and as such are key objects for testing brown dwarf
and extrasolar giant planet atmosphere models \citep{bar03}, probing
the extreme low-mass end of the initial mass function \citep{all04,me04a},
and expanding the census of the Sun's nearest neighbors.

For the past two years, we have been conducting a wide-field (74\% of the sky)
search for T dwarfs in the
Two Micron All Sky Survey \citep[hereafter 2MASS]{skr97,cut03}.  This
three-band, near-infrared ($JHK_s$) imaging survey samples the peak of the T dwarf
spectral energy distribution, and is therefore the most sensitive wide-field
sky survey currently available for identifying these cold brown dwarfs.  Our results to
date include the discovery of the bright, and therefore potentially very close
($d$ $\approx$ 8 pc) T dwarf 2MASS 1503+2525 \citep[hereafter Paper I]{me03a} and three new
T dwarfs identified in the Southern Hemisphere \citep[hereafter Paper II]{me03e}.
Here we present the discovery of seven new T dwarfs in both Northern and Southern
Hemispheres, all of which were verified by low resolution (R$\sim$150) spectroscopic
observations obtained with the IRTF 3.0m SpeX instument \citep{ray03}.

In $\S$ 2 we describe near-infrared imaging and spectroscopic observations of
T dwarf candidates and comparison stars made using SpeX and other
imaging instruments.
In $\S$ 3 we analyze these data, classifying both the new T dwarfs and
background stars, including four potential ultracool
(spectral types later than sdM7) subdwarfs,
using the low-resolution spectra and spectral comparison stars.
We also report line strengths for the 1.243/1.252 $\micron$ \ion{K}{1} doublet in
six T dwarfs observed at moderate resolution (R$\sim$1200) with SpeX, and proper
motions for all of the T dwarf discoveries.  In $\S$ 4 we discuss our results, including
distance and tangential velocity estimates, signatures of gravity and/or
metallicity in the near-infrared
spectrum of 2MASS 0034+0523, and prospects for future discoveries.
Results are summarized in $\S$ 5.

\section{Observations}

\subsection{Target Selection}

Our selection of T dwarf candidates from the 2MASS Working Point Source Database
(WPSD) is described in detail in Paper I.  In brief, we chose
point sources with $J \leq 16$,
$J-H \leq 0.3$ or $H-K_s \leq 0$, no optical counterpart with 5$\arcsec$
in the USNO A2.0 catalog \citep{mon98} or by visual inspection of Digitized Sky Survey
(DSS) images, no catalogued minor planet counterpart, and ${\mid}b{\mid} \geq 15\degr$.
Revised 2MASS photometry
for these sources is given in the All Sky Data Release
(ADR) Point Source Catalog, and is based on improved photometric calibration, particularly
at $J$-band
(see Cutri et al.\ 2003, $\S$IV.1.c).  In some cases the new photometry
pushes our candidates out of the initial selection criteria.  However, these
candidates were retained, and their ADR photometry is reported throughout this
article.  To date, a total of 912 T dwarf candidates have been observed
as part of this search program.

\subsection{Imaging}

Follow-up near-infrared imaging observations of T dwarf candidates are required
to eliminate the majority of contaminant sources.  These include
minor planets whose ephemerides were unknown or were not incorporated
at the time of 2MASS data processing,
and image artifacts that remain in the 2MASS WPSD.  Imaging observations also
provide second epoch astrometry for confirmed T dwarfs that may be used
to measure proper motion.  We therefore conducted a series of imaging campaigns
using various instrumentation on 1-4m class telescopes.

\subsubsection{IRTF 3.0m SpeX}

The SpeX instrument is comprised of a 1024$\times$1024 InSb array as
the primary detector for the spectrograph, and a second 512$\times$512
InSb array imaging/guiding camera with a $60{\arcsec}{\times}60{\arcsec}$
field-of-view (0$\farcs$12 pixels).  We made use of the latter detector to
image T dwarf candidates during
our 2003 September 17--19 (UT) IRTF observing campaign.
Conditions varied from clear to slightly hazy
with seeing between 0$\farcs$5--0$\farcs$7 at $J$-band.  Dithered exposure pairs of
30 s each were obtained at $J$-band and pair-wise subtracted for inspection.
We verified that each exposure was at least as deep as the corresponding 2MASS image.

Table 1 lists those sources that were absent
in follow-up images.  Five of these were identified as
known asteroids using the Small-Body
Search Tool maintained by the Jet Propulsion Laboratory Solar System Dynamics
Group\footnote{See \url{http://ssd.jpl.nasa.gov/cgi-bin/sb{\_}search}.}, and were typically
catalogued as asteroids after the 2MASS observation.  The remaining
sources have ecliptic latitudes ${\mid}{\beta}{\mid} < 15\degr$ and are therefore also likely to be
as yet uncataloged minor planets.

\subsubsection{Palomar 1.5m CCD Camera}

Gunn-$z$ band images of four new T dwarfs
identified in this sample (see $\S$ 3.1.2)
were obtained using the Palomar 1.5m facility CCD Camera on 2003 September 27 (UT).
Conditions were clear and photometric with 0$\farcs$7 seeing.
The CCD camera is a red-sensitive, 2048$\times$2048, thinned
detector with 24$\micron$ pixels.  Pixel scale on the sky is 0$\farcs$378.  Only
the central 1024$\times$1024 region was used for a total
field of view of 6$\farcm$5 on a side.
Raw images were bias-subtracted and divided by
a median-combined flat field frame generated from a series of lamp on/off exposures
reflected from the interior dome.  All four T dwarf targets were detected in these images
to sufficient signal-to-noise (S/N $\gtrsim$ 10) to obtain reliable astrometry.

\subsubsection{Lick 3.0m Gemini}

The Lick 3.0m Gemini instrument \citep{mcl93} is a dual 256$\times$256 HgCdTe/InSb
camera with 0$\farcs$68 pixels and a $3{\arcmin}{\times}3{\arcmin}$ field of view.
J-band images of 2MASS 1209$-$1004 were obtained using this instrument on 2003 May 12 (UT); conditions
were clear with seeing of 1$\farcs$2.  Observations were similar to those
described in Paper I, with a dithered pair of 30 s exposures obtained
and differenced to produce the final science image; no additional calibration
was required for the astrometric measurements.

\subsubsection{AAT 3.9m IRIS2}

The AAT 3.9m IRIS2 near-infrared imager and spectrograph\footnote{See
\url{http://www.aao.gov.au/iris2/index.html}.}
is a 1024$\times$1024 HgCdTe
array camera with 0$\farcs$4486 pixels and a 7$\farcm$7$\times$7$\farcm$7 field of view.
We used this camera to observe 2MASS 1231+0847, 2MASS 1828$-$4849,
and 2MASS 2331$-$4718 on 2003 June 11 and 2003 September 10 and 13 (UT).  These data are part
of an imaging program testing the use of CH$_4$ filters to efficiently identify
and classify T dwarf candidates; the program is described in detail
in Tinney et al.\ (in preparation).  Conditions during the observations were non-photometric,
with occasional cloud patches, although data were only obtained when the sky was clear.
Seeing ranged from 0$\farcs$9--2$\farcs$0.  Targets were observed in dithered sets
of three 40 s exposures in each of the CH$_4$-s and CH$_4$-l filters, which bisect the
$H$-band about the 1.6 $\micron$ CH$_4$ band.  Images were processed via the IRIS2 data reduction
pipeline, which is modeled after the UKIRT ORAC-DR pipeline\footnote{See
\url{http://www.jach.hawaii.edu/JACpublic/UKIRT/software/oracdr/index.html}.}, and
includes bad pixel masking, flat fielding, and alignment and re-sampling of the dithered
exposures to produce a final, calibrated image.

\subsection{Spectroscopy}

A total of 66 T dwarf candidates and 33 spectral comparison stars
were observed using the low-resolution
prism mode of the SpeX spectrograph primarily over two observing runs, 21--23 May 2003
and 17--19 September 2003 (UT).  A log of observations for the candidates and comparison
stars are given in Tables 2 and 3, respectively.  Conditions during the May run ranged
from light to heavy cirrus with seeing of 0$\farcs$5--0$\farcs$9 at $J$-band.
One object, 2MASS 0034+0523, was observed
on 5 September 2003 (UT) during clear conditions with similar seeing.  The prism mode of SpeX provides
0.7--2.5 $\micron$ continuous spectroscopy in a single order.  Using the 0$\farcs$5 slit,
we obtained a resolution R$\sim$150; dispersion on the chip is 20--30 {\AA} pixel$^{-1}$.
For all observations, the instrument rotator was positioned at the parallactic angle to
mitigate differential color refraction across the broad spectral band observed.
Total integration times ranged from 12 (bright M giant and supergiant stars) to
1600 s, and were typically obtained in multiple pairs of 180 s exposures dithered along
the chip for sky subtraction.  Except for a few low declination sources, the majority of
objects were observed at airmasses $\lesssim 1.5$.  In all but one case, we observed
A0 stars selected from the Henry Draper (HD) catalog shortly before or after the target observation
and at a differential airmass $\lesssim$ 0.1 for flux calibration.
Internal flat-field and Ar arc lamps were
observed immediately after the A0 star observations for instrumental calibration.

Six T dwarfs, including five of the discoveries presented here, were also observed using
the cross-dispersed, moderate-resolution SXD mode of SpeX during the May and September
runs.  These observations are summarized in Table 4.
We employed the same 0$\farcs$5 slit as the prism observations
to obtain R$\sim$1200 spectra from 0.9--2.4 $\micron$
in four orders; an additional order subtending 0.81--0.95 $\micron$ was measured for the bright
T5.5 2MASS 1503+2525.  Pixel dispersion on the chip ranged from 2.7 to 5.3 {\AA} pixel$^{-1}$.
We observed
all targets with the slit rotated to the parallactic angle in dithered pairs of 300 s each, with total
integration times of 1800--3000 s.  A0V/A0Vn HD stars and internal calibration
lamps were observed immediately after each target observation.

All data were reduced using the Spextool package \citep{cus04}.  For both the prism and SXD
datasets, science data were corrected for linearity, pair-wise subtracted, and divided by the
corresponding median-combined flat field image.  Spectral data were optimally extracted using the
default settings for aperture and background source regions, and wavelength calibration
was determined from arc lamp and sky emission lines.  Multiple spectral observations for each
source were then median-combined after scaling the spectra to match the highest signal-to-noise
observation.  Telluric and instrumental response corrections for the science data were determined
using the method outlined
in \citet{vac03}.  For the prism observations, line shape kernels were derived from the
arc lines; for the SXD observations, they were derived from the 1.005 $\micron$ \ion{H}{1} Paschen $\delta$ line
in the A0 calibrator spectra.  Adjustments were made to the telluric spectra to compensate
for differing \ion{H}{1} line strengths and velocity shifts.  Final calibration was made by
multiplying the observed target spectrum by its respective telluric correction spectrum.
For the SXD data, multiple orders for each target exposure were scaled and combined using the
corresponding low-resolution prism spectrum as a relative flux template.

\section{Analysis}

\subsection{Low Resolution Spectroscopy}

\subsubsection{Background Stars}

Examination of the low-resolution spectra indicate that the majority of these
candidate sources
are late-type M dwarfs, based on the presence of weak H$_2$O absorption at 1.4 and 1.9 $\micron$;
CO absorption at 2.3 $\micron$; TiO and VO bands at 0.7--1.0 $\micron$; and FeH, \ion{Na}{1}, and \ion{K}{1}
absorption at $J$-band.  This result is not
unexpected, as many of these sources have $J \sim 16$, implying $R-J \gtrsim$ 3--5
at the detection limit of the DSS plates \citep{rei91}, typical for late-type M dwarfs
\citep{kir99}.  The faintness of these objects also imply significant errors in their 2MASS
photometry, which explains in part their unusually blue $J-K_s$ colors.

Nevertheless, to better understand the nature of these contaminant sources, we classified
their spectra by comparison to a suite of M- and L-type spectral standards and known objects, as listed in Table 3.
A representative sample of these comparison spectra are shown in Figure 1.  For the dwarfs,
the strengthening of the 1.4 and 1.9 $\micron$ H$_2$O bands, and 0.99 and 1.2 $\micron$
FeH bands; appearance of the 1.17 and 1.25 $\micron$ \ion{K}{1} doublets;
weakening of the 0.76, 0.82, and 0.84 $\micron$ TiO bands; shift in peak flux toward 1.3 $\micron$; and
reddening of the 1.3--2.4 $\micron$ spectral energy distribution are all correlated with spectral
type.  Subdwarfs show distinctively stronger 0.99 $\micron$ FeH absorption and weaker CO and TiO
absorption than the dwarfs for equivalent H$_2$O band strengths, as well as a shift in the peak of
their spectral energy distributions to shorter wavelengths.  Giant and supergiant M stars exhibit much
deeper H$_2$O bands and stronger VO, TiO, and CO absorption, but fairly weak or absent metal hydride bands
and atomic lines.

Using
these diagnostics and the comparison spectra, we visually classified each of the observed sources
that were not identified as T dwarfs.  These classifications, which we estimate are accurate to within $\sim$0.5-1.0
subtypes, are listed in Table 2; uncertain
numerical or luminosity classifications due to low S/N data are noted by a colon.
A large percentage of the late-type
M dwarfs exhibit \ion{H}{1} Paschen $\gamma$ emission (1.09 $\micron$): 33\% (6 of 18) of the M7-M7.5 dwarfs
and 66\% (12 of 18) of the M8-M8.5 dwarfs appear to be in emission, but none of the M5-M6.5 dwarfs.
The observed emission is consistent with the high frequency ($\sim$100\%) of H$\alpha$
emission seen in the optical spectra of M7--M8 dwarfs \citep{giz00}.

While most of the background sources exhibit spectral energy distributions and absorption features similar
to the dwarf comparison stars, a handful of sources appear to be peculiar.
These include four objects that exhibit subdwarf characteristics.
As shown in Figure 2,
2MASS 0142+0523, 2MASS 1640+1231, and 2MASS 1640+2922 all appear to be similar to or somewhat later than
the sdM7.5 LSR 2036+5059 \citep{lep03a}, with strong 0.86, 0.99, and 1.2 $\micron$ FeH absorption;
weak 2.3 $\micron$ CO absorption; moderately strong 1.4 and 1.9 $\micron$ H$_2$O absorption;
and a relatively blue 1--2.5 $\micron$ spectral slope.
Another source, 2MASS 0041+3547, exhibits spectral features similar
to the L1 standard
2MASS 1439+1929 \citep{kir99}, but has stronger 0.86 and 0.99 $\micron$ FeH absorption, weaker CO,
and a somewhat bluer 1.3--2.4 $\micron$ spectral slope.
This object may be a new early-type L subdwarf,
similar to LSR 1610-0040 \citep{lep03b}.  However,
as there currently exists no classification scheme for ultracool subdwarfs at near-infrared
wavelengths, we characterize these sources as candidate subdwarfs for the time being; further
spectral analysis will be presented in a future publication.
It should be noted that only
four subdwarfs later than type M7 are currently known \citep{me03f,lep03a,lep03b,lep03c}.

Finally, we address one additional background source, 2MASS 1733+1529, classified here as a DC 10
white dwarf based on its blackbody slope and absence of \ion{H}{1} absorption lines (Figure 3).  This object
was selected because of its absence in the first generation Palomar Sky Survey (POSS-I) $R$-band plate (Figure 4, left),
although it is present on the $B$-band plate as well as the $B$-, $R$- (Figure 4, right), and $I$-band
plates of POSS-II.  The absence of this source in the POSS-I $R$-band image was independently verified
by M.\ Gray and R.\ Humphreys to a faint limit of $R \sim$ 19.5--20.0
using the Minnesota Automated Plate Scanner Catalog\footnote{MAPS; see
\url{http://aps.umn.edu/index.html}.}, and by visual examination of POSS-I prints available
at the Caltech Astrophysics Library.
2MASS 1733+1529 is not listed in
the \citet{mco99} white dwarf catalog; and its small proper motion, 0$\farcs$05$\pm$0$\farcs$04 yr$^{-1}$,
measured from Lick 3m Gemini $J$-band imaging (see $\S$ 3.3),
excludes it from the NLTT \citep{luy79b} and revised NLTT \citep{sal03} proper motion
catalogs.  As this source is relatively bright in the optical ($R$ = 16.9 in the USNO-B1.0 Catalog;
Monet et al.\ 2003), and
given the absence of any obvious photographic anomaly, it is unclear as
to why it was undetected on the POSS-I $R$-band plate.
It is possible that 2MASS 1733+1529 was obscured in this image
by an eclipsing or transiting faint source, such
as a low-mass stellar or substellar companion.
Monitoring observations are planned to
verify or place stringent constraints on this intriguing hypothesis.

\subsubsection{T Dwarf Discoveries}

Seven of the candidates listed in Table 2 exhibit clear absorption features of H$_2$O
(1.1, 1.4, and 1.9 $\micron$) and CH$_4$ (1.3, 1.6, and 2.2 $\micron$)
in their low-resolution spectra, characteristic of T dwarfs.  Finder charts for these objects
are given in Figure 5, low-resolution spectra are diagrammed in Figure 6, and spectrophotometric properties
are listed in Table 5.  These objects exhibit a broad range of
CH$_4$ band strengths and near-infrared colors ($-0.7 < J-K_s < 0.85$), encompassing early-, mid-,
and late-type T dwarf spectral morphologies \citep{me02a,geb02}.

The new T dwarfs were classified by their low-resolution spectra
following the technique of \citet{me02a},
which is based on the established MK Process of spectral classification (e.g.,
Morgan 1984).  In addition to our candidates, we observed a set of five T dwarf spectral standards:
SDSS 0423$-$0414 (T0), SDSS 1254$-$0122 (T2), 2MASS 2254+3123 (T4), 2MASS 0243$-$2453
(T6), and 2MASS 0415$-$0935 (T8).  The T2, T6, and T8 standards are those established
in \citet{me02a}, while the T0 and T4 standards were selected
as part of an expanded list of spectral standards given in \citet{me03}. To
quantify our classifications, we used revised spectral indices from the
latter reference which sample the major H$_2$O and CH$_4$ bands in the 1--2.5 $\micron$
region while avoiding strong telluric absorption regions.  These indices
are defined as:
\begin{equation}
H_2O^J = \frac{\int{F_{1.14-1.165}}}{\int{F_{1.26-1.285}}},
\end{equation}
\begin{equation}
CH_4^J = \frac{\int{F_{1.30-1.325}}}{\int{F_{1.26-1.285}}},
\end{equation}
\begin{equation}
H_2O^H = \frac{\int{F_{1.48-1.52}}}{\int{F_{1.56-1.60}}},
\end{equation}
\begin{equation}
CH_4^H = \frac{\int{F_{1.635-1.675}}}{\int{F_{1.56-1.60}}},
\end{equation}
and
\begin{equation}
CH_4^K = \frac{\int{F_{2.215-2.255}}}{\int{F_{2.08-2.12}}},
\end{equation}
where $\int{F_{{\lambda}_1-{\lambda}_2}}$ designates the integrated flux
between wavelengths ${\lambda}_1$ and ${\lambda}_2$, and indices are defined as the ratio of
flux at the base of the absorption band to the nearby pseudo-continuum\footnote{The presence
of overlying opacity throughout the 1--2.5 $\micron$ region implies that no true continuum
is present; we therefore normalize to the local spectral maximum, or pseudo-continuum.}.
These spectral indices were measured for all T dwarfs in our low-resolution spectral sample.
Classifications for each index (excluding the spectral standards) were determined as the closest
match to the standard values, allowing for subtypes halfway between the standard classes
(i.e., integer subclasses).
Final classifications for each object were determined as the mean of the individual
index classifications, rounded off to the nearest 0.5 subclass.

Table 6 lists the spectral indices and derived classifications for the new
and previously known T dwarfs.
The scatter amongst individual index subtypes is typically 0.4--0.7 subclasses, with some objects
exhibiting no scatter, justifying our 0.5 subclass precision.  As a check, we compared derived
subtypes to those from the literature for seven previously identified and classified T dwarfs;
all were consistent within 0.5 subclasses.  We also examined the behavior
of the indices with spectral type, as shown in Figure 7; all five indices show minimal
scatter about a line connecting the standard values, again consistent with the adopted
classification precision.  The scatter in these indices is in fact better than that
seen in the classifications of \citet{me02a} and \citet{geb02}, reflecting overall higher
signal-to-noise data and the improved spectral index definitions.
The spectral types for the T dwarf discoveries
range from T3 (2MASS 1209$-$1004) to T7 (2MASS 0034+0523), the former object being the earliest-type
T dwarf so far identified in our 2MASS search.  As shown in Figure 6, the derived classifications
are consistent with a monotonic increase in H$_2$O and CH$_4$ bandstrengths with spectral type.

\subsection{Moderate Resolution Spectroscopy}

Moderate resolution SpeX spectra of six T dwarfs are shown in Figure 8.
These data reveal the complex H$_2$O and CH$_4$ molecular features in far greater detail than the prism
spectra.  While the S/N of the higher resolution spectra are on average lower (10--50\%),
most of the structure seen is real and repeats between the objects.  These data also resolve
the 1.243/1.252 $\micron$ \ion{K}{1} doublet, shown in detail in Figure 8,
a key diagnostic of temperature at the brightest peak
in the spectral energy distribution.
Pseudo-equivalent widths (PEWs; equivalent width relative to the pseudo-continuum)
for these lines were measured following the method outlined in Paper II, directly integrating over
the line profiles. Results are
given in Table 7.  The strongest lines are found in the T5 2MASS 2331$-$4718, which have PEWs of
8.5$\pm$0.7 and 12.8$\pm$0.6 {\AA} for the 1.243 and 1.252 $\micron$ \ion{K}{1} lines, respectively.  Indeed,
with the exception of the T5.5 2MASS 0516$-$0445 (for which only low S/N data have
been obtained), these are the strongest \ion{K}{1} lines measured in
any T dwarf to date \citep{me02a,me03e,mcl03}.  Both 2MASS 1503+2525 and 2MASS 1231+0847 exhibit
broadened \ion{K}{1} lines, possibly due to rapid rotation or high photospheric pressure, the latter case
indicative of high surface gravity.  Also note
the nearly absent \ion{K}{1} lines in the T7 2MASS 0034+0523, due either to its low T$_{eff}$
or possibly metal deficiency (see $\S$ 4.2).  Overall, the observed line strengths are in
general agreement
with previous work, with the strongest \ion{K}{1} lines found amongst the T5 dwarfs and becoming
progressively weaker toward the later spectral types.

\subsection{Astrometry}

Proper motions for the T dwarf discoveries were measured using 2MASS ADR data and
the follow-up imaging observations described in $\S$ 2.2.  Data analysis was similar to that described in
Paper II.  We used 2MASS catalog data as first epoch astrometry, while the follow-up images,
taken roughly 3--5 years after the 2MASS observations, comprised our second epoch dataset.
2MASS sources within the re-imaged areas (excluding the T dwarf)
were matched to detected sources on the follow-up images.  First-order coordinate solutions for the images were
then determined by linear regression using this grid of background stars,
allowing for the rejection of 3$\sigma$ outliers (i.e., moving sources).
For the IRIS2 (CH$_4$-s images only)
and CCD observations, $\sim$15--60 background sources were
used, yielding positional uncertainties of 0$\farcs$1--0$\farcs$4, equivalent to 2MASS astrometric
accuracy \citep{cut03}, and proper motion uncertainties of $\lesssim$ 0$\farcs$1 yr$^{-1}$.
For the Gemini observation of 2MASS 1209$-$1004, only four background objects were available
and astrometric uncertainties were assumed to be somewhat higher.
Using these coordinate solutions, the
second epoch position of the T dwarf was computed and its motion derived.
We note that the agreement between proper motion determinations for
2MASS 0034+0523 obtained with the Palomar 60''
CCD Camera and AAT IRIS2 lend confidence to the reliability
of our measurements.

Table 8 lists the
resulting proper motions.  As expected for a nearby population of faint
brown dwarfs, these objects have large motions in general, with 2MASS 1231+0847
exhibiting the largest at 1$\farcs$55$\pm$0$\farcs$07 yr$^{-1}$.  This object would
be easily identified in a near-infrared proper motion survey.  On the other hand, two
T dwarfs, 2MASS 0407+1514 and 2MASS 2331$-$4718, have motions below our sensitivity limits.
We discuss the associated tangential velocities for these T dwarfs below.

\section{Discussion}

\subsection{Spectrophotometric Distances}

Using the derived classifications and 2MASS photometry, it is possible to estimate
the spectrophotometric distances of our T dwarf discoveries. We employed the polynomial
$J$- and $K_s$-band absolute magnitude/spectral type relations from
\citet{tin03}, based on parallax measurements from their own program and from
\citet{dah02}.  Distance estimates for each of the newly discovered T dwarfs
were calculated in both bands (with the exception of
2MASS 0034+0523 which was not detected at $K_s$ by 2MASS) and for $\pm$0.5 subclasses about the nominal
classification.  Final distances and uncertainties were determined by the mean and standard deviation
of these estimates, respectively, and are listed in Table 5.  Assuming that they are single
sources, all of the objects listed are within 25
pc of the Sun, with the most distant object being the earliest-type T dwarf 2MASS 1209$-$1004.
Three objects, 2MASS 0034+0523, 2MASS 1231+0847, and 2MASS 1828$-$4849, are at or within
10 pc from the Sun within the reported uncertainties,
with the first (and latest-type) object having an estimated distance of only 8.2$\pm$1.5 pc.  It should
be noted that the uncertainties in these distance estimates do not take into account systematic deviations
in the absolute magnitude/spectral type relation \citep{tin03} nor possible duplicity, and should be confirmed by
parallax measurement.

Combining the distance estimates with the proper motion determinations from $\S$ 3.3, we calculated
tangential velocities for these T dwarfs, listed in Table 8.  The mean $V_{tan}$ of those T dwarfs
with detected motion is 43 km s$^{-1}$ with a standard deviation of 28 km s$^{-1}$; including the upper limits
yields a somewhat lower mean of 33 km s$^{-1}$.  This value is consistent
with the mean $V_{tan}$ for disk dwarfs (39 km s$^{-1}$; Reid \& Hawley 2000) but is somewhat higher
than that for field late-type M and L dwarfs (22 km s$^{-1}$; Gizis et al.\ 2000), suggesting that the
T dwarfs in this sample may be drawn from a somewhat older population.
A similar difference in the $V_{tan}$ distribution between field L and T dwarfs has
also been noted by \citet{vrb04}, and a mean age difference between these classes is
predicted in field substellar mass function simulations \citep{me04a}.
This possible age segregation is consistent with the evolution
of brown dwarfs, as an object of a given mass will evolve from warm (L dwarf) to cold (T dwarf)
as it ages; however, a larger sample
must be considered before drawing any firm conclusions about the relative ages
of the L and T dwarf field populations.

We note that the spectrophotometric distance of 2MASS 1231+0847 is consistent within its uncertainties
to the nearby (13.4$\pm$0.3 pc; HIPPARCHOS; Perryman et al.\ 1997) K7V star Gliese 471, located roughly
8$\farcm$1 (6500 AU) to the northwest. Indeed, 2MASS 1231+0847
was also identified in a parallel search for wide brown dwarf companions to nearby stars currently being
conducted by J.\ D.\ Kirkpatrick.  However, while their direction of motion is nearly identical ($\sim$230$\degr$),
2MASS 1231+0847 has a proper motion nearly twice as large as Gliese 471, and is therefore not a bound
companion.

\subsection{Gravity/Metallicity Signatures at K-band}

The T7 2MASS 0034+0523 has the bluest $J-K_s$ color amongst the T dwarf discoveries,
and it may be even bluer as 2MASS photometry provides only an upper limit.
Examination of near-infrared spectral data confirms this
color, as 2MASS 0034+0523 exhibits a fairly suppressed
$K$-band peak in comparison to the rest of the T6--T8 dwarfs observed.  This is quantified in Figure 9,
which plots the logarithm of the spectral ratio
\begin{equation}
K/J = \frac{\int{F_{2.06-2.10}}}{\int{F_{1.25-1.29}}}
\end{equation}
as a function of spectral type for all of the T dwarfs
observed.  SpeX prism data are ideally suited for this measurement, as the
full near-infrared spectral range is sampled in a single order and no correction is required to match
the relative scalings between the $J$- and $K$-bands.
The ratio exhibits a tight linear trend across the full spectral type range of T dwarfs, although
there is a somewhat greater spread in values amongst the T5--T6 dwarfs.  2MASS 0034+0523 stands well above
this trend, however, having the smallest value of $K/J$, and hence the most suppressed $K$-band peak,
in the entire sample.

The spectral properties of this source are reminiscent of the
peculiar T6 2MASS 0937+2931 \citep{me02a}, which not only has a very
blue near-infrared color ($J-K_s = -0.62{\pm}0.14$; Paper I), likely due to enhanced
CIA H$_2$ absorption, but also strong
absorption from the pressure-broadened 0.77 $\micron$ \ion{K}{1} resonance doublet and
the 0.99 $\micron$ FeH band \citep{me03d}.  The enhanced pressure-sensitive features are symptomatic of
a high pressure photosphere, which can exist on a brown dwarf with a high surface
gravity (i.e., old and massive) and/or a metal deficient atmosphere \citep{bur02}.
Indeed, strong FeH and CIA H$_2$ absorption are hallmarks of
cool halo subdwarf spectra (Figure 1),
which are themselves typically older and metal-poor,
and 2MASS 0937+2931 has been interpreted as a possible
thick disk or halo brown dwarf \citep{bur02,me03d}.  2MASS 0034+0523, like
2MASS 0937+2931, also has exceedingly weak 1.243/1.252 $\micron$ \ion{K}{1} lines that may be due to
reduced metallicity or higher surface gravity \citep{bur02},
although its late spectral type (and hence cool temperature) may be the dominant factor.
2MASS 0034+0523 does not have a large $V_{tan}$ as might be expected for a halo star,
although on an individual basis this does not rule out its membership in an older
kinematic population.  Clearly, a more detailed study
of both of these peculiar T dwarfs is needed to assess metallicity and/or gravity effects
in cool brown dwarf spectra.

\subsection{Prospects for Further Discoveries}

To date, we have observed roughly 70\% of our 2MASS search sample, identifying
31 T dwarfs, 8 of which have estimated or measured \citep{dah02,tin03,vrb04}
distances within 10 pc of the Sun.  This is roughly consistent with the
predicted numbers from Paper I ($\sim$35--45 T dwarfs), although we have uncovered
less than half of the $\sim$20 T dwarfs predicted to have distances less than
10 pc.  Many of the latter are probably late-type T dwarfs,
T$_{eff}$ $\lesssim$ 1000 K, too faint to be detected by 2MASS beyond a few parsecs.
In addition to these very low-luminosity sources, our search has identified
one {\em bona-fide} ultracool halo subdwarf, the late-type sdL 2MASS 0532+8246 \citep{me03f},
and now four candidate late-type subdwarfs requiring further verification.
In retrospect, the search criteria employed are
well-suited for identifying these metal-deficient objects, which
have red optical/near-infrared colors, peak in flux at $J$-band,
and exhibit relatively blue $J-K_s$ colors
due to enhanced CIA H$_2$ absorption \citep{leg00}.
Such serendipitous discoveries
provide a new opportunity for exploring the physical properties, particularly metallicity and age diagnostics,
of very cool stars and brown dwarfs.

\section{Summary}

We have discovered seven new T dwarfs in the 2MASS survey with spectral
types ranging from T3 to T7, spectrophotometric distances of 8.2 to 23 pc, and proper motions
from our detection limit (0$\farcs$1 yr$^{-1}$) to 1$\farcs$55 yr$^{-1}$.  This sample
adds substantially to the current census of T dwarfs,
now around 50 objects.  We have also
identified four candidate ultracool subdwarfs, including one possible early-type L subdwarf, which share the
blue near-infrared colors and red optical/near-infrared colors of our T dwarf discoveries.
We estimate that several T dwarfs remain to be identified in our sample, and possibly many more
cool subdwarfs, yielding new targets in our quest to understand the observational properties of the
lowest-mass stars and brown dwarfs in the Solar Neighborhood.

\acknowledgments

We thank our telescope operators Bill Golisch, Dave Griep, and Paul Sears,
and instrument specialist John Rayner, for their
support during the IRTF observations, and the IRTF TAC for its generous allocation
of time for this project.  We also thank
Matt Gray and Roberta Humphreys at the
University of Minnesota, Jean Mueller at Palomar Observatory,
and the library staff at the
Caltech Astrophysics Library,
for their assistance
in obtaining, examining, and verifying
POSS-I images of 2MASS 1733+1529.
We are grateful to our referee for her/his prompt and thorough review
of the original manuscript.
A.\ J.\ B.\ acknowledges support provided by NASA through
Hubble Fellowship grant HST-HF-01137.01 awarded by the Space Telescope Science Institute,
which is operated by the Association of Universities for Research in Astronomy,
Incorporated, under NASA contract NAS5-26555.
K.\ L.\ C.\ acknowledges support from a NSF Graduate Research Fellowship.
This research has made use of
the SIMBAD database, operated at CDS, Strasbourg, France.
POSS-I, POSS-II, SERC, and AAO $R$-band images were obtained from the Digitized Sky Survey
image server maintained by the Canadian Astronomy Data Centre,
which is operated by the Herzberg Institute of Astrophysics,
National Research Council of Canada. The Digitized Sky Survey was produced at the
Space Telescope Science Institute under U.S. Government grant NAG W-2166.
The images of these surveys are based on photographic data obtained using the
Oschin Schmidt Telescope on Palomar Mountain and the UK Schmidt Telescope.
The plates were processed into the present compressed digital form with the
permission of these institutions.
This publication makes use of data from the Two
Micron All Sky Survey, which is a joint project of the University
of Massachusetts and the Infrared Processing and Analysis Center,
funded by the National Aeronautics and Space Administration and
the National Science Foundation.
2MASS data were obtained from
the NASA/IPAC Infrared Science Archive, which is operated by the
Jet Propulsion Laboratory, California Institute of Technology,
under contract with the National Aeronautics and Space
Administration.
The authors wish to extend special thanks to those of Hawaiian ancestry
on whose sacred mountain we are privileged to be guests.
Electronic copies of the spectra presented here can be obtained directly from
the primary author.

\clearpage

\begin{deluxetable}{lrlcccl}
\tabletypesize{\scriptsize}
\tablecaption{T Dwarf Candidates Absent in Follow-up SpeX Imaging.}
\tablewidth{0pt}
\tablehead{
 & & \multicolumn{4}{c}{2MASS Observations\tablenotemark{b}} & \\
\cline{3-6}
\colhead{Object\tablenotemark{a}} &
\colhead{$\beta$ ($\degr$)} &
\colhead{UT Date} &
\colhead{$J$} &
\colhead{$J-H$} &
\colhead{$H-K_s$} &
\colhead{Identification\tablenotemark{c}} \\
\colhead{(1)} &
\colhead{(2)} &
\colhead{(3)} &
\colhead{(4)} &
\colhead{(5)} &
\colhead{(6)} &
\colhead{(7)} \\}
\startdata
2MASS J00344717$-$3343404 & $-$34 & 1998 Oct 3 & 15.77$\pm$0.05 & 0.40$\pm$0.10 & $-$0.17$\pm$0.22 &  1999 XJ197  \\
2MASS J02132776+1939028 & 6 & 1997 Oct 19 & 15.87$\pm$0.06 & 0.30$\pm$0.13 & 0.18$\pm$0.19 &   \\
2MASS J02273983+1824524 & 4 & 1997 Oct 20 & 15.25$\pm$0.05 & 0.52$\pm$0.07 & $-$0.10$\pm$0.12 & 2001 TL10  \\
2MASS J02322515+2942172 & 14 & 1997 Nov 10 & 14.61$\pm$0.04 & 0.30$\pm$0.05 & 0.18$\pm$0.06 &   \\
2MASS J02381005+2725209 & 11 & 1997 Nov 10 & 15.56$\pm$0.05 & 0.36$\pm$0.09 & $-$0.11$\pm$0.15 &   \\
2MASS J02470189+2200236 & 6 & 1997 Oct 15 & 15.73$\pm$0.07 & 0.15$\pm$0.15 & 0.41$\pm$0.22 &   \\
2MASS J03153063+2101434 & 3 & 2000 Nov 8 & 15.87$\pm$0.09 & 0.10$\pm$0.18 & $-$0.07$\pm$0.27 &   \\
2MASS J03483407+3231433 & 12 & 1998 Jan 25 & 15.54$\pm$0.05 & 0.35$\pm$0.10 & $-$0.01$\pm$0.15 & 2002 AW2  \\
2MASS J04122822+1316535 & $-$8 & 2000 Nov 26 & 15.96$\pm$0.07 & 0.64$\pm$0.12 & 0.00$\pm$0.17 & 2000 UA110  \\
2MASS J04185615+2527216 & 4 & 1997 Nov 29 & 15.96$\pm$0.06 & 0.33$\pm$0.11 & $-$0.01$\pm$0.19 &   \\
2MASS J04265381+1158063 & $-$10 & 1998 Nov 18 & 15.90$\pm$0.08 & 0.08$\pm$0.18 & 0.38$\pm$0.22 & 2001 JU3  \\
\enddata
\tablenotetext{a}{All objects are listed with their 2MASS ADR Point Source Catalog source
designations, given as ``2MASS Jhhmmss[.]ss$\pm$ddmmss[.]s''. The
suffix is the
sexagesimal Right Ascension and declination at J2000 equinox.}
\tablenotetext{b}{Photometry from the 2MASS ADR; note that some objects have revised photometry
placing them outside of our original WPSD search constraints.}
\tablenotetext{c}{Asteroid identifications are from the Small-Body
Search Tool maintained by the Jet Propulsion Laboratory Solar System Dynamics Group:
\url{http://ssd.jpl.nasa.gov/cgi-bin/sb{\_}search}.}
\end{deluxetable}

\begin{deluxetable}{lcccccclcl}
\rotate
\tabletypesize{\scriptsize}
\tablecaption{Log of SpeX Prism Observations: T Dwarf Candidates.}
\tablewidth{0pt}
\tablehead{
\colhead{Object} &
\colhead{$J$\tablenotemark{a}} &
\colhead{$J-H$\tablenotemark{a}} &
\colhead{$H-K_s$\tablenotemark{a}} &
\colhead{UT Date} &
\colhead{t (s)} &
\colhead{Airmass} &
\colhead{Calibrator} &
\colhead{SpT} &
\colhead{Identification\tablenotemark{b}} \\
\colhead{(1)} &
\colhead{(2)} &
\colhead{(3)} &
\colhead{(4)} &
\colhead{(5)} &
\colhead{(6)} &
\colhead{(7)} &
\colhead{(8)} &
\colhead{(9)} &
\colhead{(10)} \\}
\startdata
2MASS J00013044+1010146 & 15.83$\pm$0.07 & 0.69$\pm$0.12 & $-$0.07$\pm$0.07 & 2003 Sep 19 & 720  & 1.02  & HD 210501  & A0 V &  M6V \\
2MASS J00115060$-$1523450 & 15.93$\pm$0.08 & 0.29$\pm$0.17 & 0.38$\pm$0.08 & 2003 Sep 18 & 720  & 1.29  & HD 219833  & A0 V &  M7.5Ve \\
2MASS J00335534$-$0908247 & 15.96$\pm$0.09 & 0.96$\pm$0.14 & $-$0.24$\pm$0.09 & 2003 Sep 19 & 720  & 1.15  & HD 1154  & A0 V & M8Ve:  \\
2MASS J00345157+0523050 & 15.54$\pm$0.05 & 0.09$\pm$0.09 &  $<$ $-$0.8 & 2003 Sep 05 & 1800  & 1.05  & HD 5267  & A1 Vn &  T7 \\
2MASS J00412179+3547133 & 15.94$\pm$0.08 & 0.21$\pm$0.17 & 0.56$\pm$0.08 & 2003 Sep 19 & 720  & 1.08  & HD 13869  & A0 V &  sdL? \\
2MASS J00552554+4130184 & 15.81$\pm$0.08 & 0.84$\pm$0.11 & $-$0.02$\pm$0.08 & 2003 Sep 18 & 720  & 1.16  & HD 7215  & A0 V &  M8Ve: \\
2MASS J00583814$-$1747311 & 15.94$\pm$0.09 & 0.20$\pm$0.19 & 0.27$\pm$0.09 & 2003 Sep 18 &  720  & 1.28  & HD 219833  & A0 V &  M6V \\
2MASS J01045111$-$3327380 & 15.96$\pm$0.10 & 0.19$\pm$0.19 & 0.94$\pm$0.10 & 2003 Sep 18 & 720  & 1.72  & HD 12275  & A0 V &  M8V \\
2MASS J01151621+3130061  & 15.89$\pm$0.10 & 0.22$\pm$0.20 & 0.66$\pm$0.24 & 2003 Sep 19 & 720  & 1.05  & HD 13869  & A0 V &  M8.5Ve \\
2MASS J01423153+0523285 & 15.91$\pm$0.08 & 0.32$\pm$0.14 & $-$0.01$\pm$0.08 & 2003 Sep 17 & 720  & 1.05  & HD 18571  & A0 V &  sdM7.5? \\
2MASS J01470204+2120242 & 15.99$\pm$0.07 & 0.97$\pm$0.21 & $-$0.01$\pm$0.07 & 2003 Sep 18 & 720  & 1.09  & HD 7215  & A0 V &  M7.5Ve: \\
2MASS J01532750+3631482 & 15.81$\pm$0.06 & 0.66$\pm$0.11 & 0.44$\pm$0.06 & 2003 Sep 18 & 720  & 1.10  & HD 7215  & A0 V &  M6V \\
2MASS J03023398$-$1028223 & 16.24$\pm$0.12 & 0.52$\pm$0.19 & 0.60$\pm$0.12 & 2003 Sep 17 & 720  & 1.17  & HD 25792  & A0 V &  M5.5V \\
2MASS J04035944+1520502 & 15.97$\pm$0.08 & 0.29$\pm$0.16 & 0.42$\pm$0.08 & 2003 Sep 19 & 720  & 1.03  & HD 25175  & A0 V &  M7V \\
2MASS J04070885+1514565 & 16.06$\pm$0.09 & 0.04$\pm$0.23 & 0.10$\pm$0.09 & 2003 Sep 19 & 1080  & 1.01  & HD 25175  & A0 V &  T5.5 \\
2MASS J04071296+1710474 & 16.01$\pm$0.10 & 0.32$\pm$0.19 & 0.54$\pm$0.10 & 2003 Sep 19 &  720 & 1.00  & HD 25175  & A0 V &  M8V \\
2MASS J04360273+1547536 & 16.13$\pm$0.09 & 0.98$\pm$0.12 & 0.22$\pm$0.09 & 2003 Sep 18 & 720  & 1.00  & HD 25175  & A0 V &  M6.5V \\
2MASS J11070582+2827226 & 15.74$\pm$0.06 & 0.57$\pm$0.12 & 0.11$\pm$0.06 & 2003 May 23 & 1080  & 1.01  & HD 89239  & A0 V &  M8Ve \\
2MASS J11150577+2520467 & 15.85$\pm$0.08 & 0.82$\pm$0.11 & $-$0.04$\pm$0.08 & 2003 May 23 & 1080  & 1.01  & HD 89239  & A0 V &  M7.5Ve: \\
2MASS J11323833$-$1446374 & 15.83$\pm$0.09 & 0.20$\pm$0.14 & 0.29$\pm$0.09 & 2003 May 21 & 1200  & 1.00 &  HD 101369   & A0 V &  M7V \\
2MASS J11463232+0203414 & 15.89$\pm$0.09 & 0.21$\pm$0.19 & 0.21$\pm$0.09 & 2003 May 22 & 1200  & 1.05  & HD 97585  & A0 V &  M5.5V \\
2MASS J12095613$-$1004008 & 15.91$\pm$0.07 & 0.58$\pm$0.12 & 0.27$\pm$0.07 & 2003 May 22 & 1440  & 1.02  & HD 109309  & A0 V &  T3 \\
 &  &  &  & 2003 May 23 & 1080  & 1.16  & HD 109309  & A0 V &  \\
2MASS J12121714$-$2253451 & 15.69$\pm$0.07 & 0.28$\pm$0.16 & 0.37$\pm$0.07 & 2003 May 23 & 720  & 1.39  & HD 110649  & A0 V &  M8Ve \\
2MASS J12314753+0847331 & 15.57$\pm$0.07 & 0.26$\pm$0.13 & 0.09$\pm$0.07 & 2003 May 21 & 1080  & 1.02  & HD 111744  & A0 V &  T6 \\
2MASS J12575768$-$0204085 & 15.71$\pm$0.17 & 0.46$\pm$0.25 & 0.48$\pm$0.17 & 2003 May 22 & 1080  & 1.11  & HD 109309  & A0 V &  sd:M5 \\
2MASS J13272391+0946446 & 15.99$\pm$0.10 & 0.65$\pm$0.15 & $-$0.06$\pm$0.10 & 2003 May 23 & 720  & 1.02  & HD 116960  & A0 V &  M6V \\
2MASS J13593574+3031039 & 15.87$\pm$0.08 & 0.26$\pm$0.14 & 0.34$\pm$0.08 & 2003 May 21 & 800  & 1.05  & HD 121626  & A0 &  M7V \\
2MASS J14171672$-$0407311 & 15.95$\pm$0.07 & 0.59$\pm$0.10 & $-$0.12$\pm$0.07 & 2003 May 22 & 1080  & 1.10  & HD 126129  & A0 V &  M8Ve: \\
2MASS J15243203+0934386 & 15.05$\pm$0.05 & 0.79$\pm$0.08 & $-$0.04$\pm$0.05 & 2003 May 22 & 720  & 1.04  & HD 136831  & A0 V &  M7Ve: \\
2MASS J15412408+5425598 & 15.93$\pm$0.08 & 0.26$\pm$0.18 & 0.32$\pm$0.08 & 2003 May 21 & 1080  & 1.22  & HD 142282  & A0 &  M8Ve \\
2MASS J15561873+1300527 & 15.91$\pm$0.07 & 0.13$\pm$0.17 & 0.93$\pm$0.07 & 2003 May 23 & 720  & 1.01  & HD 140729  & A0 V &  M8.5Ve \\
2MASS J15590462$-$0356280 & 15.97$\pm$0.07 & 0.25$\pm$0.17 &  $<$ 0.3 & 2003 May 21 &  720 & 1.10  & HD 143396  & A0 &  M8.5Ve \\
2MASS J16304206$-$0232224 & 15.49$\pm$0.06 & 0.51$\pm$0.10 & 0.23$\pm$0.06 & 2003 May 22 & 720  & 1.09  & HD 145647  & A0 V &  M8V \\
2MASS J16390818+2839015 & 15.85$\pm$0.08 & 0.50$\pm$0.13 & 0.48$\pm$0.08 & 2003 May 23 & 720  & 1.02  & HD 145647  & A0 V &  M7.5V \\
2MASS J16403197+1231068  & 15.95$\pm$0.08 & 0.34$\pm$0.14 & $<$ 0.4 & 2003 May 21 & 720  & 1.01  & HD 151545  & A0 &  sdM8? \\
2MASS J16403561+2922225 & 15.70$\pm$0.07 & 0.63$\pm$0.10 & 0.40$\pm$0.07 & 2003 May 22 & 720  & 1.07  & HD 145647  & A0 V & sdM8? \\
2MASS J17252029$-$0024508 & 15.91$\pm$0.08 & 0.87$\pm$0.11 & $-$0.03$\pm$0.08 & 2003 May 23 & 720  & 1.14  & HD 171149  & A0 V &  M5V \\
2MASS J17330480+0041270 & 15.91$\pm$0.10 & 0.13$\pm$0.19 & 0.37$\pm$0.10 & 2003 May 21 & 720  & 1.06  & HD 159835  & A0 &  M7V \\
2MASS J17331764+1529116 & 15.91$\pm$0.07 & 0.45$\pm$0.14 &  $<$ 0.0 & 2003 May 21 & 720  & 1.01  & HD 159907  & A0 &  DC 10 \\
2MASS J17364839+0220426 & 15.85$\pm$0.08 & 0.49$\pm$0.13 & 0.47$\pm$0.08 & 2003 May 23 & 720  & 1.13  & HD 171149  & A0 V &  M8V \\
2MASS J18112466+3748513 & 15.54$\pm$0.05 & 0.32$\pm$0.08 & $-$0.26$\pm$0.05 & 2003 May 22 & 720  & 1.07  & HD 174567  & A0 V &  mid M:: \\
2MASS J18244344+2937133 & 15.89$\pm$0.08 & 0.08$\pm$0.15 &  $<$ 0.1 & 2003 May 23 & 720  & 1.08  & HD 165029  & A0 V &  M6V \\
2MASS J18283572$-$4849046 & 15.18$\pm$0.06 & 0.27$\pm$0.09 & $-$0.27$\pm$0.06 & 2003 Sep 18 & 1080  & 2.7-2.8  & HD 177406    & A0 V &  T6 \\
2MASS J18411320$-$4000124 & 15.94$\pm$0.06 & 0.92$\pm$0.11 & $-$0.20$\pm$0.06 & 2003 Sep 17 & 720  & 2.07  & HD 176425  & A0 V &  M7.5Ve: \\
2MASS J18530004$-$4133275 & 15.68$\pm$0.07 & 0.62$\pm$0.11 & $-$0.05$\pm$0.07 & 2003 Sep 18 & 720  & 2.10  & HD 176425  & A0 V &  M7 pec \\
2MASS J19010601+4718136 & 15.86$\pm$0.07 & 0.39$\pm$0.12 & $-$0.17$\pm$0.07 & 2003 May 21 & 1440  & 1.13  & HD 177390  & A0 &  T5 \\
2MASS J19240765$-$2239504 & 15.97$\pm$0.09 & 0.24$\pm$0.17 & 0.40$\pm$0.09 & 2003 May 21 & 1080  & 1.37  & HD 174072  & A0 V &  M6V \\
2MASS J19312708+5948588 & 15.47$\pm$0.07 & 0.59$\pm$0.11 & 0.20$\pm$0.07 & 2003 Sep 18 & 720  & 1.35  & HD 194354  & A0 Vs &  M5.5V \\
2MASS J19445221$-$0831036 & 15.82$\pm$0.06 & 0.50$\pm$0.10 & $-$0.08$\pm$0.06 & 2003 Sep 17 & 720   & 1.18  & HD 188489   & A0 V &  M6V \\
2MASS J19522109$-$5059169 & 15.96$\pm$0.08 & 0.30$\pm$0.19 & 0.76$\pm$0.08 & 2003 Sep 18 & 720  & 3.04  &  HD 183626 & A0 V &  M6V \\
2MASS J19570817$-$1627558 & 15.84$\pm$0.09 & 0.29$\pm$0.17 & 0.77$\pm$0.09 & 2003 Sep 19 & 720  & 1.32  & HD 198787  & A0 V &  M6V \\
2MASS J20494090+1140068 & 16.26$\pm$0.10 & 0.62$\pm$0.18 & 0.69$\pm$0.10 & 2003 Sep 17 & 720  & 1.02  & HD 210501  & A0 V &  M7.5V \\
2MASS J21100889+2132483 & 15.94$\pm$0.08 & 0.66$\pm$0.11 & 0.000$\pm$0.08 & 2003 Sep 17 &  720 & 1.01  & HD 210501  & A0 V &  M8V \\
2MASS J21105305+1903568 & 15.94$\pm$0.08 & 0.17$\pm$0.14 & 0.71$\pm$0.08 & 2003 Sep 19 & 720  & 1.02  & HD 210501  & A0 V &  M8.5Ve \\
2MASS J21214516+2825375 & 15.81$\pm$0.08 & 0.28$\pm$0.15 & 0.74$\pm$0.08 & 2003 Sep 19 & 720  & 1.06  & HD 210501  & A0 V &  M7.5V \\
2MASS J21353463+2352085 & 16.10$\pm$0.12 & 0.67$\pm$0.16 & 0.05$\pm$0.12 & 2003 Sep 19 &  720 & 1.06  & HD 210501  & A0 V &  M7V \\
2MASS J22270083$-$1231482 & 15.68$\pm$0.06 & 0.49$\pm$0.10 & $-$0.13$\pm$0.06 & 2003 Sep 17 & 720  & 1.18  & HD 219833  & A0 V &  M5.5V \\
2MASS J22425680+0720249 & 16.05$\pm$0.10 & 0.88$\pm$0.15 & 0.14$\pm$0.10 & 2003 Sep 18 & 720  & 1.03  & HD 210501  & A0 V &  M8Ve: \\
2MASS J22453832$-$0722060 & 16.11$\pm$0.09 & 0.55$\pm$0.14 & 0.18$\pm$0.09 & 2003 Sep 17 & 720  & 1.12  & HD 219833  & A0 &  M7 pec \\
2MASS J22465014$-$0643357 & 15.50$\pm$0.06 & 0.53$\pm$0.11 & 0.29$\pm$0.06 & 2003 Sep 17 & 720  & 1.12  & HD 219833  & A0 &  M5V \\
2MASS J23270985+2341364 & 15.99$\pm$0.10 & 0.24$\pm$0.19 & 0.35$\pm$0.10 & 2003 Sep 17 & 720  & 1.10  & HD 222749  & A0 V &  M7.5Ve: \\
2MASS J23312378$-$4718274 & 15.66$\pm$0.07 & 0.15$\pm$0.16 & 0.12$\pm$0.07 & 2003 Sep 17 & 720  & 2.7-2.8  & HD 216009  & A0 V &  T5 \\
2MASS J23354680+1257273 & 15.83$\pm$0.06 & 0.12$\pm$0.15 & 0.70$\pm$0.06 & 2003 Sep 18 & 720  & 1.02  & HD 210501  & A0 V &  M8.5Ve \\
2MASS J23363834+4523306 & 15.99$\pm$0.10 & 0.61$\pm$0.15 & $-$0.27$\pm$0.10 & 2003 Sep 17 & 720  & 1.26  & HD 222749 & A0 V &  M8V \\
2MASS J23480816+4052343 & 16.11$\pm$0.09 & 0.46$\pm$0.16 & 0.54$\pm$0.09 & 2003 Sep 17 & 720  & 1.23  & HD 222749  & A0 V &  M7.5V \\
2MASS J23592315$-$1239079 & 15.96$\pm$0.07 & 0.28$\pm$0.15 & 0.72$\pm$0.07 & 2003 Sep 18 & 720  & 1.22  & HD 219833  & A0 V &  M7V \\
\enddata
\tablenotetext{a}{Photometry from the 2MASS ADR; note that some objects have revised photometry
outside of our original WPSD search constraints.}
\tablenotetext{b}{Classifications are based on comparison to spectral templates (Table 3) and
for M dwarfs are accurate to within 0.5-1.0 subclasses (see $\S$ 3.1.1).
Classification terms followed by a colon (``:'') are more uncertain due to poor S/N data
or lack of adequate comparison stars.}
\end{deluxetable}

\begin{deluxetable}{lcccccclcll}
\rotate
\tabletypesize{\tiny}
\tablecaption{Log of SpeX Prism Observations: Comparison Stars.}
\tablewidth{0pt}
\tablehead{
\colhead{Object} &
\colhead{$J$\tablenotemark{a}} &
\colhead{$J-H$\tablenotemark{a}} &
\colhead{$H-K_s$\tablenotemark{a}} &
\colhead{UT Date} &
\colhead{t (s)} &
\colhead{Airmass} &
\colhead{Calibrator} &
\colhead{SpT} &
\colhead{Identification} &
\colhead{Ref} \\
\colhead{(1)} &
\colhead{(2)} &
\colhead{(3)} &
\colhead{(4)} &
\colhead{(5)} &
\colhead{(6)} &
\colhead{(7)} &
\colhead{(8)} &
\colhead{(9)} &
\colhead{(10)} &
\colhead{(11)} \\}
\startdata
2MASS J01514155+1244300 & 16.57$\pm$0.13 & 0.96$\pm$0.17 & 0.42$\pm$0.13 & 2003 Sep 19 & 1080  & 1.03  & HD 13869  & A0 V &  T1:  & 1 \\
2MASS J02431371$-$2453298 & 15.38$\pm$0.05 & 0.24$\pm$0.12 & $-$0.08$\pm$0.05 & 2003 Sep 17 & 1080  & 1.42  & HD 27616  & A0 V &  T6  & 2 \\
2MASS J04151954$-$0935066 & 15.70$\pm$0.06 & 0.16$\pm$0.13 & 0.11$\pm$0.06 & 2003 Sep 17 & 1080  & 1.15  & HD 25792  & A0 V &  T8  & 2 \\
2MASS J04234858$-$0414035 & 14.47$\pm$0.03 & 1.00$\pm$0.04 & 0.53$\pm$0.03 & 2003 Sep 17 & 720  & 1.10  & HD 31411  & A0 V &  T0  & 1 \\
2MASS J11040127+1959217 & 14.38$\pm$0.03 & 0.90$\pm$0.04 & 0.53$\pm$0.03 & 2003 May 21 & 960  & 1.02  & HD 101060  & A0 V &  L4 & 3 \\
2MASS J11240487+3808054 & 12.71$\pm$0.02 & 0.70$\pm$0.04 & 0.45$\pm$0.02 & 2003 May 22 & 360  & 1.05  & HD 98152  & A0 V &  M8.5V & 3 \\
2MASS J12255432$-$2739466AB & 15.26$\pm$0.05 & 0.16$\pm$0.09 & 0.02$\pm$0.05 & 2003 May 23 & 720  & 1.55  & HD 110141  & A0 V &  T6/T8:  & 4,5 \\
2MASS J12545393$-$0122474 & 14.89$\pm$0.04 & 0.80$\pm$0.04 & 0.25$\pm$0.04 & 2003 May 22 & 720  & 1.08  & HD 109309  & A0 V &  T2  & 6 \\
2MASS J14392836+1929149 & 12.76$\pm$0.02 & 0.72$\pm$0.03 & 0.50$\pm$0.02 & 2003 May 23 & 360  & 1.00  & HD 122945  & A0 V &  L1 & 7 \\
2MASS J14571496$-$2121477 & 15.32$\pm$0.05 & 0.06$\pm$0.10 & 0.03$\pm$0.05 & 2003 May 22 & 1080  & 1.33  & HD 133466  & A0 V &  Gl 570D, T8  & 8 \\
2MASS J15031961+2525196 & 13.94$\pm$0.02 & 0.08$\pm$0.04 & $-$0.11$\pm$0.02 & 2003 May 22 & 720  & 1.02  & HD 136831  & A0 V &  T5.5  & 9 \\
2MASS J15261405+2043414 & 15.59$\pm$0.06 & 1.09$\pm$0.07 & 0.57$\pm$0.06 & 2003 May 23 & 720  & 1.05  & HD 140729  & A0 &  L7 & 10 \\
2MASS J16241436+0029158 & 15.49$\pm$0.05 & $-$0.03$\pm$0.11 &  $<$ 0.0 & 2003 May 22 & 1440  & 1.09  & HD 145647  & A0 V &  T6  & 11 \\
2MASS J16322911+1904407 & 15.87$\pm$0.07 & 1.25$\pm$0.08 & 0.61$\pm$0.07 & 2003 May 22 & 1080  & 1.07  & HD 145647  & A0 V &  L8  & 7 \\
2MASS J17072343$-$0558249AB & 12.05$\pm$0.02 & 0.79$\pm$0.04 & 0.55$\pm$0.02 & 2003 May 23 & 960  & 1.19  & HD 171149  & A0 V &  M9V/L3 & 12,13 \\
2MASS J17361766+1346225 & 10.41$\pm$0.02 & 0.60$\pm$0.03 & 0.28$\pm$0.02 & 2003 May 21 & 180  & 1.05  & HD 165029  & A0 V &  LP 508-14, M4V & 14 \\
2MASS J17503293+1759042 & 16.34$\pm$0.10 & 0.39$\pm$0.17 & 0.47$\pm$0.10 & 2003 May 23 & 1080  & 1.08  & HD 165029  & A0 V &  T3.5 & 1 \\
2MASS J18261131+3014201 & 11.66$\pm$0.02 & 0.48$\pm$0.03 & 0.36$\pm$0.02 & 2003 May 21 & 240  & 1.18  & HD 174567  & A0 Vs &  M8.5V & 15 \\
2MASS J19165762+0509021 & 9.91$\pm$0.03 & 0.68$\pm$0.04 & 0.46$\pm$0.03 & 2003 Sep 19 & 200  & 1.04  & HD 189920  & A0 V &  VB10, M8V & 16 \\
2MASS J20282035+0052265 & 14.30$\pm$0.04 & 0.92$\pm$0.05 & 0.58$\pm$0.04 & 2003 May 23 & 720  & 1.06  & HD 198070  & A0 Vn &  L3  & 17 \\
2MASS J20362186+5059503 & 13.61$\pm$0.03 & 0.45$\pm$0.05 & 0.22$\pm$0.05 & 2003 Sep 18 & 720  & 1.18  & HD 194354  & A0 Vs &  LSR 2036+50; sdM7.5 & 15,18 \\
2MASS J20392378$-$2926335 & 11.36$\pm$0.03 & 0.61$\pm$0.03 & 0.38$\pm$0.03 & 2003 May 22 & 720  & 1.57  & HD 202941  & A0 V &  M6V & 19 \\
2MASS J20491972$-$1944324 & 12.85$\pm$0.02 & 0.63$\pm$0.03 & 0.44$\pm$0.02 & 2003 Sep 19 & 360  & 1.34  & HD 198787  & A0 V &  M7.5V & 19 \\
2MASS J20575409$-$0252302 & 13.12$\pm$0.02 & 0.85$\pm$0.03 & 0.54$\pm$0.02 & 2003 May 23 & 360  & 1.09  & HD 198070  & A0 Vn &  L1.5 & 20 \\
2MASS J21073169$-$0307337 & 14.20$\pm$0.03 & 0.76$\pm$0.04 & 0.56$\pm$0.03 & 2003 May 23 & 720  & 1.10  & HD 198070  & A0 Vn &  M9V & 3, 21 \\
2MASS J21225635+3656002 & 13.71$\pm$0.03 & 0.41$\pm$0.05 & 0.18$\pm$0.05 & 2003 Sep 18 & 720  & 1.05  & HD 209932  & A0 V &  LSR 2122+36; esdM5 & 15,18 \\
2MASS J22120345+1641093 &  11.43$\pm$0.03 & 0.60$\pm$0.04 & 0.28$\pm$0.03 & 2003 Sep 19 & 240  & 1.03  & HD 210501  & A0 V &  M5V & 14 \\
2MASS J22282889$-$4310262 & 15.66$\pm$0.07 & 0.30$\pm$0.14 & 0.07$\pm$0.07 & 2003 Sep 17 & 1080  & 2.3-2.4  & HD 216009  & A0 V &  T6.5 & 22 \\
2MASS J22341394+2359559 &  13.15$\pm$0.02 & 0.79$\pm$0.03 & 0.52$\pm$0.03 & 2003 Sep 19 & 360  & 1.02  & HD 210501  & A0 V &  M9.5V & 19 \\
2MASS J22541892+3123498 & 15.26$\pm$0.05 & 0.24$\pm$0.09 & 0.12$\pm$0.05 & 2003 Sep 18 & 720  & 1.04  & HD 209932  & A0 V &  T4  & 2 \\
V* V1451 Aql & 2.3$\pm$0.3 & 0.9$\pm$0.4 & 0.3$\pm$0.4 & 2003 May 22 & 12 & 1.16 & HD 185533 & A0 V & M5III\tablenotemark{b} & 23 \\
V* Z Cep & 5.95$\pm$0.03 & 0.88$\pm$0.05 & 0.67$\pm$0.05 & 2003 May 22 & 72 & 1.27 & HD 203893 & A0 V & MIab\tablenotemark{b} & 23 \\
SV* P 2312 & 1.9$\pm$0.3 & 0.9$\pm$0.4 & 0.4$\pm$0.4 & 2003 May 22 & 12 & 1.28 & HD 203893 & A0 V & M7III\tablenotemark{b} & 23 \\
\enddata
\tablenotetext{a}{Photometry from the 2MASS ADR.}
\tablenotetext{b}{Spectral types for variable giant stars may vary by several subclasses over time, and
given classifications may not accurately represent the actual spectrum observed.}
\tablerefs{(1) \citet{geb02}; (2) \citet{me02a}; (3) \citet{cru03}; (4) \citet{me99};
(5) \citet{me03c}; (6) \citet{leg00}; (7) \citet{kir99}; (8) \citet{me00a}; (9) Paper I;
(10) \citet{kir00}; (11) \citet{str99}; (12) \citet{giz02}; (13) McElwain \& Burgasser, in prep.;
(14) J.\ Gizis, priv.\ comm.; (15) \citet{lep02}; (16) \citet{kir91}; (17) \citet{haw02}; (18) \citet{lep03a};
(19) \citet{giz00}; (20) Kirkpatrick \& Tinney, in prep.; (21) Cruz et al., in prep.; (22) Paper II; (23) SIMBAD}
\end{deluxetable}

\begin{deluxetable}{llccclc}
\tabletypesize{\scriptsize}
\tablecaption{Log of SpeX SXD Observations.}
\tablewidth{0pt}
\tablehead{
\colhead{Object} &
\colhead{SpT} &
\colhead{UT Date} &
\colhead{t (s)} &
\colhead{Airmass} &
\colhead{Calibrator} &
\colhead{SpT} \\
\colhead{(1)} &
\colhead{(2)} &
\colhead{(3)} &
\colhead{(4)} &
\colhead{(5)} &
\colhead{(6)} &
\colhead{(7)} \\}
\startdata
2MASS J00345157+0523050 &  T7 & 2003 Sep 19 & 3000  & 1.04  & HD 6457  & A0 Vn  \\
2MASS J12314753+0847331  & T6 & 2003 May 21 & 3000  & 1.08  & HD 111744  & A0 V  \\
2MASS J15031961+2525196 &  T5.5 & 2003 May 23 & 1800  & 1.01  & HD 131951  & A0 V  \\
2MASS J18283572$-$4849046  & T6 & 2003 Sep 19 & 3000  & 2.7-2.8  & HD 177406  & A0 V  \\
2MASS J19010601+4718136 & T5 & 2003 Sep 19 & 2400  & 1.19  & HD 178207  & A0 Vn  \\
2MASS J23312378$-$4718274  & T5 & 2003 Sep 18 & 2400  & 2.5-2.6  & HD 216009  & A0 V \\
\enddata
\end{deluxetable}

\begin{deluxetable}{llccccc}
\tabletypesize{\scriptsize}
\tablecaption{Spectral and Photometric Properties of 2MASS T Dwarf Discoveries.}
\tablewidth{0pt}
\tablehead{
\colhead{Name} &
\colhead{SpT} &
\colhead{2MASS $J$} &
\colhead{$J-H$} &
\colhead{$H-K_s$} &
\colhead{$d$\tablenotemark{a}} \\
\colhead{} &
\colhead{} &
\colhead{} &
\colhead{} &
\colhead{} &
\colhead{(pc)} \\
\colhead{(1)} &
\colhead{(2)} &
\colhead{(3)} &
\colhead{(4)} &
\colhead{(5)} &
\colhead{(6)} \\
}
\startdata
2MASS 0034+0523 & T7   & 15.54$\pm$0.05 & 0.09$\pm$0.09 & $<$ $-$0.8 & 8.2$\pm$1.4  \\
2MASS 0407+1514 & T5.5   & 16.06$\pm$0.09 & 0.04$\pm$0.23 & 0.10$\pm$0.33 &  19$\pm$3  \\
2MASS 1209$-$1004 & T3   & 15.91$\pm$0.07 & 0.58$\pm$0.12 & 0.27$\pm$0.17 &  23.2$\pm$2.1  \\
2MASS 1231+0847 & T6  & 15.57$\pm$0.07 & 0.26$\pm$0.13 & 0.09$\pm$0.22 & 12.0$\pm$2.4  \\
2MASS 1828$-$4849 & T6   & 15.18$\pm$0.06 & 0.27$\pm$0.09 & $-$0.27$\pm$0.16 & 10.6$\pm$2.1  \\
2MASS 1901+4718 & T5 &  15.86$\pm$0.07 & 0.39$\pm$0.12 & $-$0.17$\pm$0.30 &  20$\pm$3  \\
2MASS 2331$-$4718 & T5   & 15.66$\pm$0.07 & 0.15$\pm$0.16 & 0.12$\pm$0.25 & 18.0$\pm$2.5  \\
\enddata
\tablenotetext{a}{Spectrophotometric distance estimate; see $\S$ 4.1.}
\end{deluxetable}

\begin{deluxetable}{llcccccl}
\tabletypesize{\small}
\tablecaption{Classification Indices for Observed T Dwarfs.}
\tablewidth{0pt}
\tablehead{
 & \colhead{Prior} & \multicolumn{5}{c}{Indices\tablenotemark{a}} & \colhead{Derived} \\
\colhead{Name} &
\colhead{SpT\tablenotemark{b}} &
\colhead{H$_2$O$^J$} &
\colhead{CH$_4$$^J$} &
\colhead{H$_2$O$^H$} &
\colhead{CH$_4$$^H$} &
\colhead{CH$_4$$^K$} &
\colhead{SpT} \\
\colhead{(1)} &
\colhead{(2)} &
\colhead{(3)} &
\colhead{(4)} &
\colhead{(5)} &
\colhead{(6)} &
\colhead{(7)} &
\colhead{(8)} \\
}
\startdata
\cline{1-8}
\multicolumn{8}{c}{Standards} \\
\cline{1-8}
SDSS 0423$-$0414 & T0 & 0.62 & 0.95 & 0.64 & 1.07 & 0.83 & \\
SDSS 1254$-$0122 & T2 & 0.46 & 0.91 & 0.48 & 1.00 & 0.59 & \\
2MASS 2254+3123 & T4 & 0.36 & 0.87 & 0.39 & 0.63 & 0.31 & \\
2MASS 0243$-$2453 & T6 & 0.14 & 0.66 & 0.30 & 0.36 & 0.16 & \\
2MASS 0415$-$0935 & T8 & 0.04 & 0.44 & 0.18 & 0.11 & 0.05 & \\
\cline{1-8}
\multicolumn{8}{c}{Known T Dwarfs} \\
\cline{1-8}
SDSS 0151+1244 & T1$\pm$1 & 0.64(0) & 0.95(0) & 0.64(0) & 1.03(1) & 0.70(1) & T0.5 \\
SDSS 1750+1759 & T3.5 & 0.46(2) & 0.87(4) & 0.45(3) & 0.71(4) & 0.35(4) & T3.5 \\
2MASS 1503+2525 & T5.5 & 0.24(5) & 0.74(5) & 0.34(5) & 0.42(6) & 0.20(5) & T5 \\
2MASS 1225$-$2739AB & T6 & 0.17(6) & 0.67(6) & 0.28(6) & 0.35(6) & 0.18(6) & T6 \\
SDSS 1624+0029 & T6 & 0.16(6) & 0.71(6) & 0.31(6) & 0.34(6) & 0.14(6) & T6 \\
2MASS 2228$-$4310 & T6.5 & 0.15(6) & 0.68(6) & 0.29(6) & 0.28(7) & 0.12(7) & T6.5 \\
Gliese 570D & T8 & 0.06(8) & 0.50(8) & 0.20(8) & 0.15(8) & 0.10(8) & T8 \\
\cline{1-8}
\multicolumn{8}{c}{Discoveries} \\
\cline{1-8}
2MASS 1209$-$1004 &  & 0.40(3) & 0.83(4) & 0.45(3) & 0.77(3) & 0.64(3) & T3 \\
2MASS 1901+4718 &  & 0.28(5) & 0.78(5) & 0.36(5) & 0.47(5) & 0.24(5) & T5 \\
2MASS 2331$-$4718 &  & 0.19(6) & 0.72(5) & 0.33(5) & 0.49(5) & 0.21(5) & T5 \\
2MASS 0407+1514 &  & 0.23(5) & 0.76(5) & 0.34(5) & 0.40(6) & 0.16(6) & T5.5 \\
2MASS 1828$-$4849 &  & 0.18(6) & 0.70(6) & 0.31(6) & 0.40(6) & 0.20(5) & T6 \\
2MASS 1231+0847\tablenotemark{c} &  & 0.18(6) & 0.68(6) & 0.27(6) & 0.39(6) & 0.17(6) & T6 \\
2MASS 0034+0523 &  & 0.10(7) & 0.64(6) & 0.23(8) & 0.25(7) & 0.13(7) & T7 \\
\enddata
\tablenotetext{a}{Indices are defined in \citet{me03}. Values in parentheses for
each index for known and new T dwarfs are the
closest spectral type match to the standard indices; see \citet{me02a}.}
\tablenotetext{b}{Assigned spectral types for spectral standards \citep{me02a,me03};
published spectral types for known T dwarfs \citep[objects in common have identical
spectral types]{me02a,geb02}.}
\tablenotetext{c}{Based on combined spectral data over two nights.}
\end{deluxetable}

\begin{deluxetable}{llccccc}
\tabletypesize{\small}
\tablecaption{\ion{K}{1} Pseudo-equivalent Widths.}
\tablewidth{0pt}
\tablehead{
 & &
\multicolumn{2}{c}{1.243 $\micron$} & &
\multicolumn{2}{c}{1.252 $\micron$} \\
\cline{3-4} \cline{6-7}
\colhead{Object} &
\colhead{SpT} &
\colhead{$\lambda$$_c$ ($\micron$)} &
\colhead{pEW ({\AA})} & &
\colhead{$\lambda$$_c$ ($\micron$)} &
\colhead{pEW ({\AA})} \\
\colhead{(1)} &
\colhead{(2)} &
\colhead{(3)} &
\colhead{(4)} &  &
\colhead{(5)} &
\colhead{(6)} \\}
\startdata
2MASS 1901+4718 & T5 & 1.243 & 6.6$\pm$0.8 & & 1.252 & 8.9$\pm$0.9 \\
2MASS 2331$-$4718 & T5 & 1.244 & 8.5$\pm$0.7 & & 1.252 & 12.8$\pm$0.6  \\
2MASS 1503+2525 & T5.5 & 1.243 & 4.8$\pm$0.3 & & 1.252 & 7.4$\pm$0.4  \\
2MASS 1828$-$4849 & T6 & 1.243 & 3.9$\pm$0.7 & & 1.252 & 6.6$\pm$0.7  \\
2MASS 1231+0847 & T6 & 1.243 & 4.9$\pm$0.6 & & 1.253 & 6.7$\pm$0.6  \\
2MASS 0034+0523 & T7 & 1.241 & 1.7$\pm$0.5 & & 1.252 & 3.9$\pm$0.4 \\
\enddata
\end{deluxetable}

\begin{deluxetable}{llcccccc}
\tabletypesize{\footnotesize}
\tablecaption{Kinematics of 2MASS T Dwarf Discoveries.}
\tablewidth{0pt}
\tablehead{
\colhead{Name} &
\colhead{SpT} &
\colhead{$\mu$} &
\colhead{$\theta$} &
\colhead{$V_{tan}$} &
\colhead{$\Delta$T} &
\colhead{No.\ Stars} &
\colhead{Source\tablenotemark{a}} \\
\colhead{} &
\colhead{} &
\colhead{($\arcsec$ yr$^{-1}$)} &
\colhead{($\degr$)} &
\colhead{(km s$^{-1}$)} &
\colhead{(yr)} &
\colhead{Used in Fit} \\
\colhead{(1)} &
\colhead{(2)} &
\colhead{(3)} &
\colhead{(4)} &
\colhead{(5)} &
\colhead{(6)} &
\colhead{(7)} &
\colhead{(8)} \\
}
\startdata
2MASS 0034+0523 & T7 &  0.68$\pm$0.06 & 72$\pm$4 &  26$\pm$5 & 3.07 & 17 & O \\
 &  &  0.70$\pm$0.14 & 74$\pm$13 &  27$\pm$7 & 3.02 & 21 & A \\
2MASS 0407+1514 & T5.5  & $<$ 0.09 & ... & $<$ 8 & 5.19 & 13 & O \\
2MASS 1209$-$1004 & T3 & 0.46$\pm$0.10 & 140$\pm$8 & 51$\pm$12 & 4.26 & 4 & G \\
2MASS 1231+0847 & T6 &  1.55$\pm$0.07 & 228$\pm$3 &  88$\pm$18 & 3.24 & 20 & A \\
2MASS 1828$-$4849 & T6 &  0.33$\pm$0.07 & 50$\pm$12  & 17$\pm$5 & 2.92 & 39 & A \\
2MASS 1901+4718 & T5 &  0.38$\pm$0.02 & 197$\pm$3 & 35$\pm$6 & 5.27 & 62 & O \\
2MASS 2331$-$4718 & T5 & $<$ 0.10  & ...  & $<$ 9 & 2.90 & 13 & A \\
\enddata
\tablenotetext{a}{O = Palomar 1.5m CCD $z$-band imaging; G = Lick 3.0m Gemini $J$-band
imaging; A = AAT 3.9m IRIS2 CH$_4$-s imaging.}
\end{deluxetable}

\clearpage

\onecolumn

\begin{figure}
\centering
\epsscale{1.1}
\plottwo{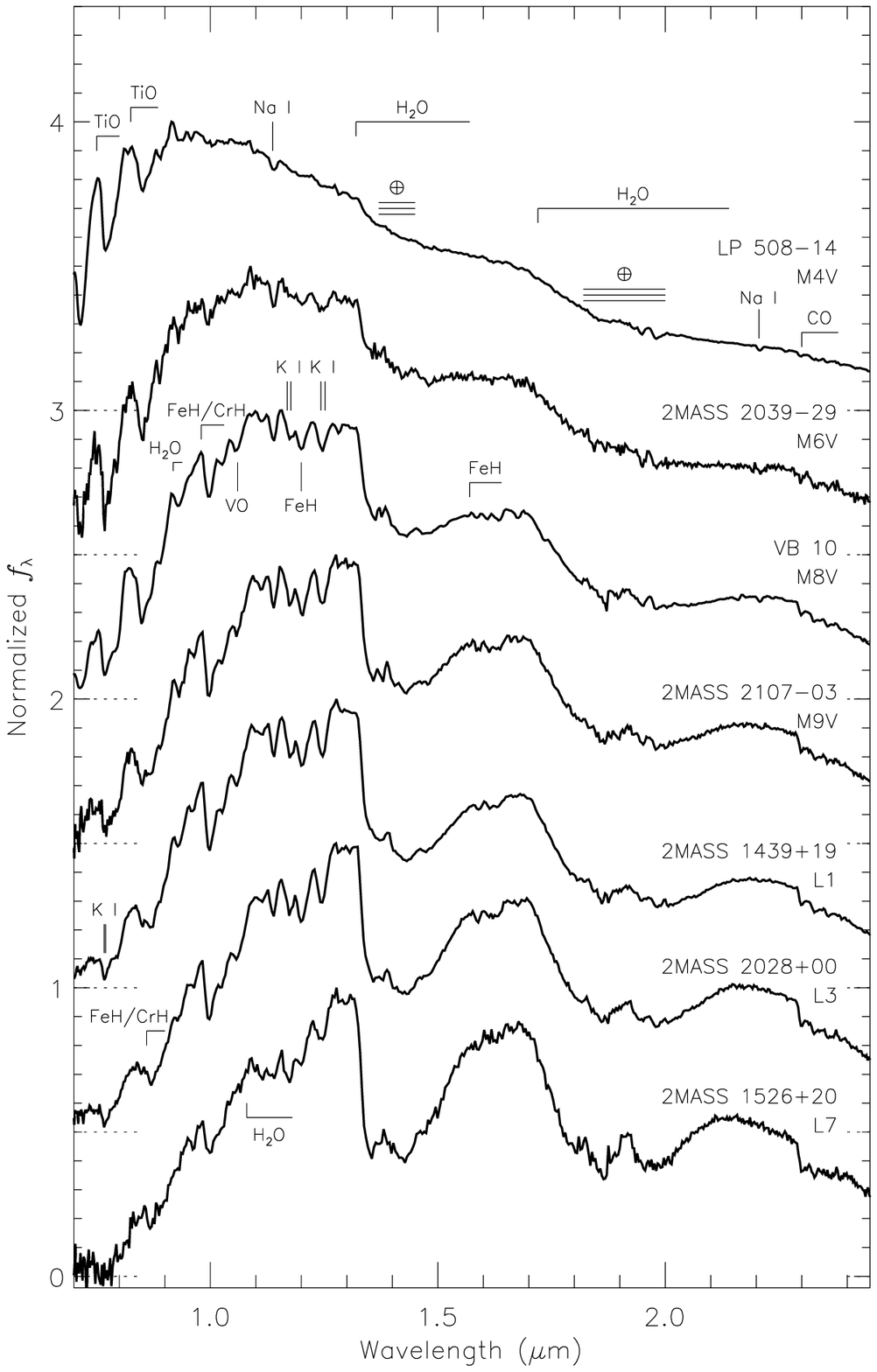}{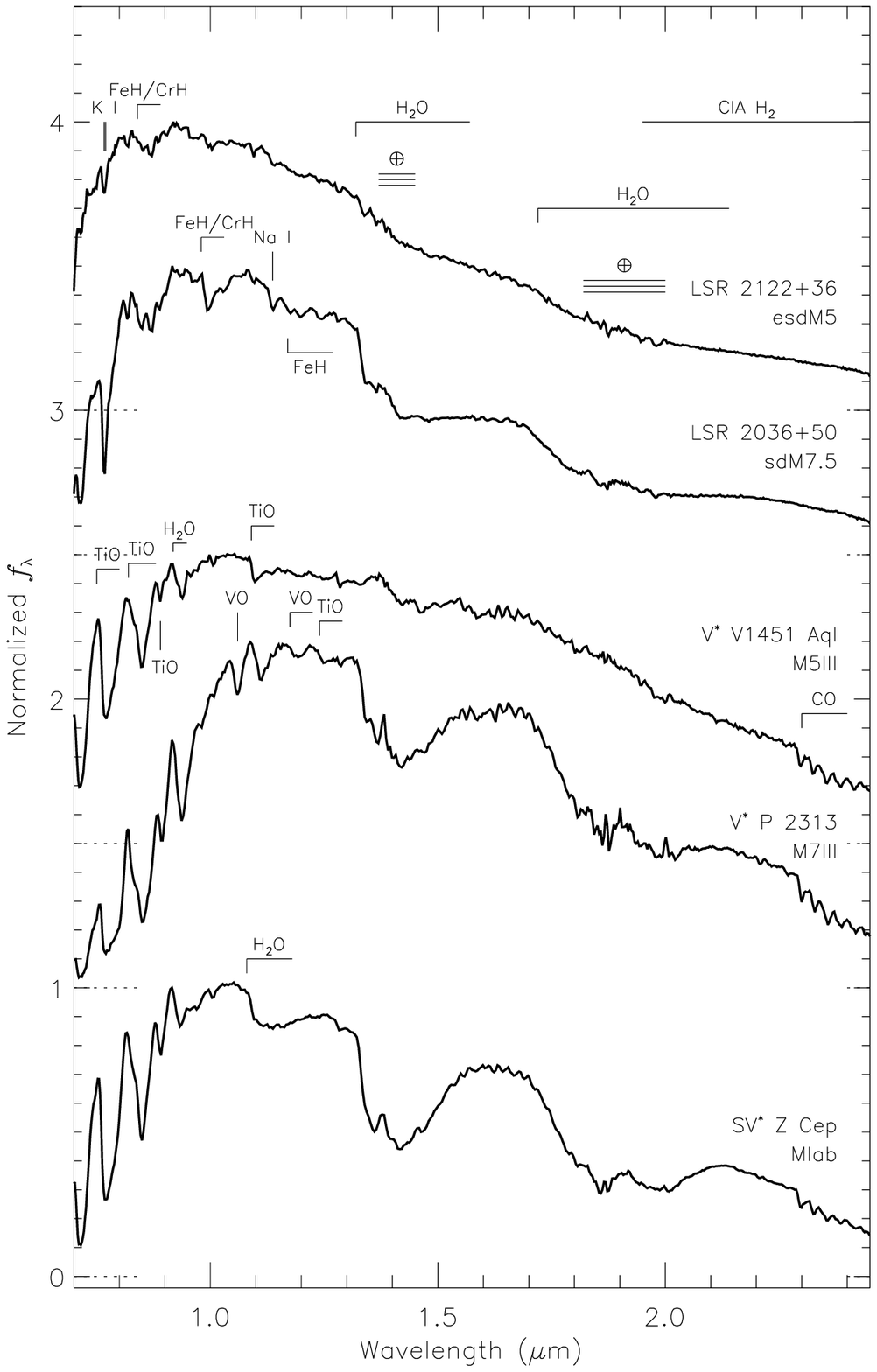}
\caption{SpeX prism spectra for comparison stars.  ({\em Left}) M and L dwarfs,
including the optical spectral standards VB 10 (M8; Kirkpatrick, Henry, \& McCarthy 1991) and
2MASS 1439+1929 (L1; Kirkpatrick et al.\ 1999).
({\em Right}) M-type subdwarfs, giants,
and supergiants.  Key spectral features
are noted in both plots, as well as regions of strong telluric absorption ($\earth$).
All spectra are in units of F$_{\lambda}$
normalized at their peak flux density and offset by a constant (dotted lines).}
\end{figure}

\begin{figure}
\epsscale{0.8}
\plotone{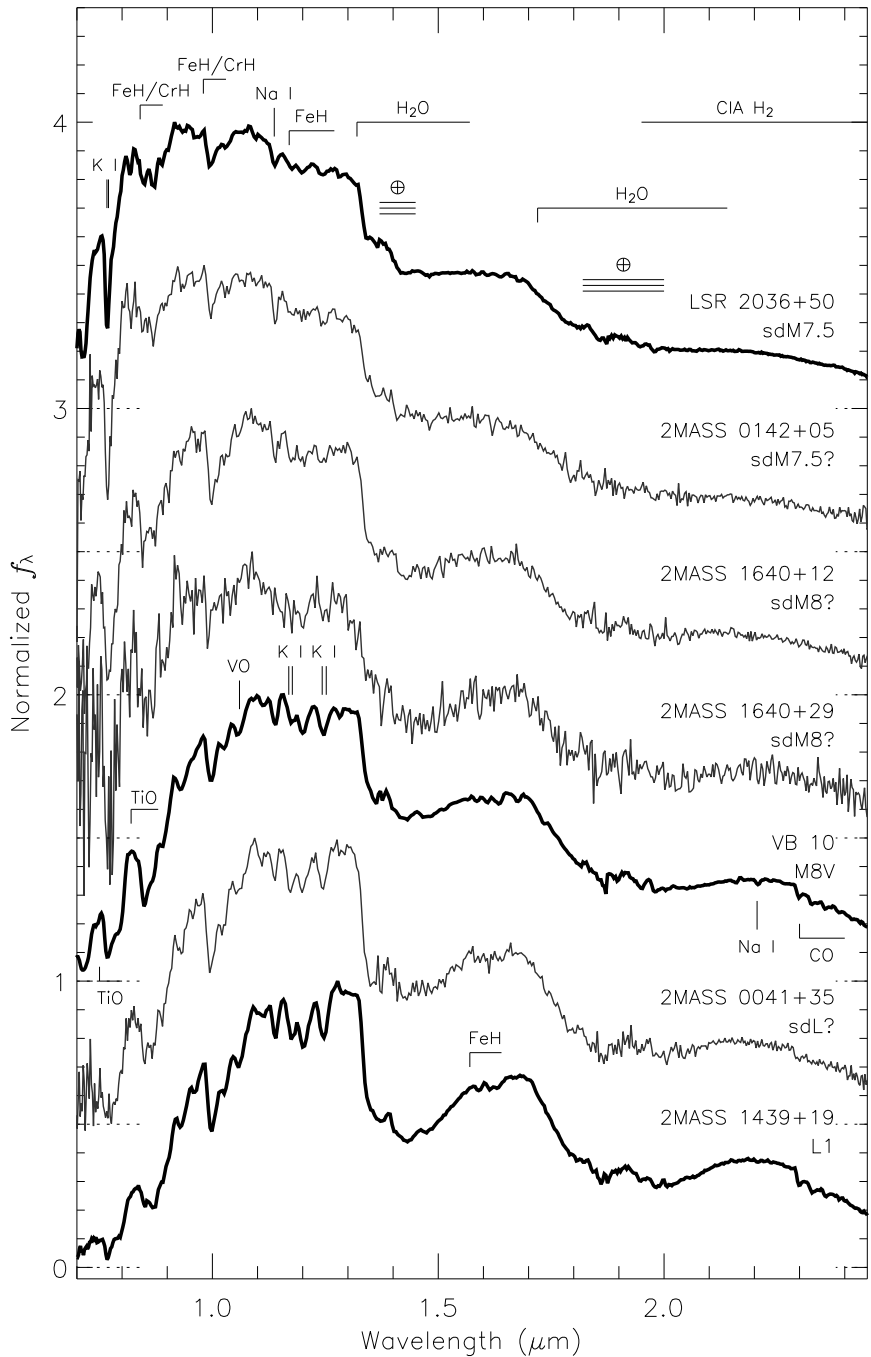}
\caption{SpeX prism spectra of candidate subdwarfs (thin grey lines)
identified in this sample.  Spectra are normalized
at their flux density peak and offset by a constant (dotted lines). Comparison stars
LSR 2036+5059 (sdM7.5), VB 10 (M8V), and 2MASS 1439+1929 (L1) are plotted with thick black lines.
Features are noted as in Figure 1.}
\end{figure}

\begin{figure}
\epsscale{0.5}
\centering
\plotone{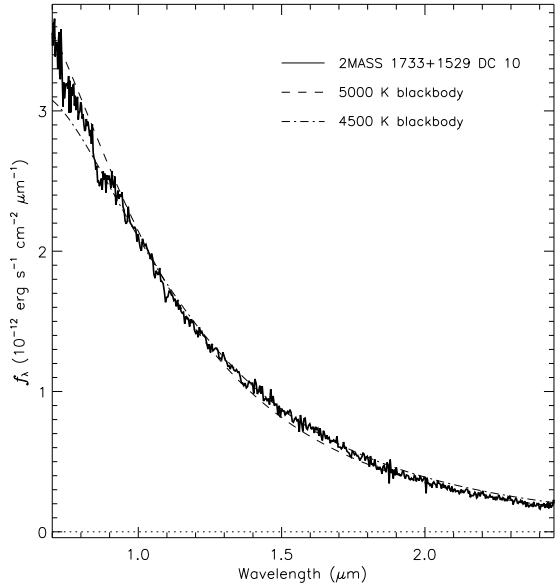}
\caption{SpeX prism spectrum of 2MASS 1733+1529 (solid line).
Best-fit blackbody curves of 5000 K (dashed line)
and 4500 K (dot-dashed line) are shown.}
\end{figure}

\begin{figure}
\epsscale{0.5}
\centering
\plotone{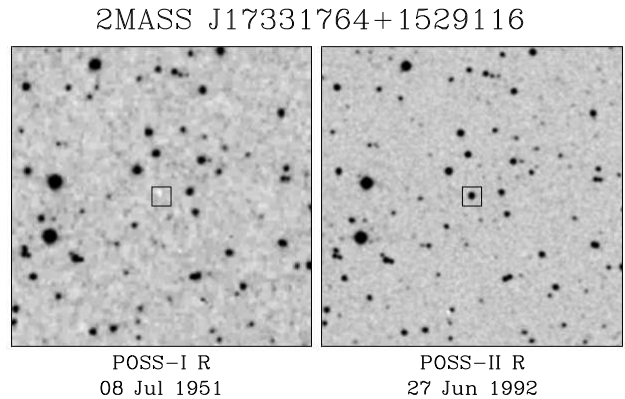}
\caption{POSS-I (left) and POSS-II (right) $R$-band images of the 2MASS 1733+1529 field.
Images are scaled to the same spatial resolution, 5$\arcmin$ on a side, with North up and East
to the left.  The small proper motion
measured for this object ($0{\farcs}5{\pm}0{\farcs}4$ yr$^{-1}$) implies that its absence on the POSS-I
$R$-band plate is not due to motion.}
\end{figure}

\begin{figure}
\centering
\caption{Finder charts for the T dwarf discoveries, showing
second generation $R$-band DSS (POSS-II, SERC, AAO)
and 2MASS ($J$- and $K_s$-bands) fields.
Images are scaled and oriented as in Figure 4. [Note that these are now in gif format attached as separate files] }
\end{figure}

\begin{figure}
\epsscale{1.1}
\centering
\plottwo{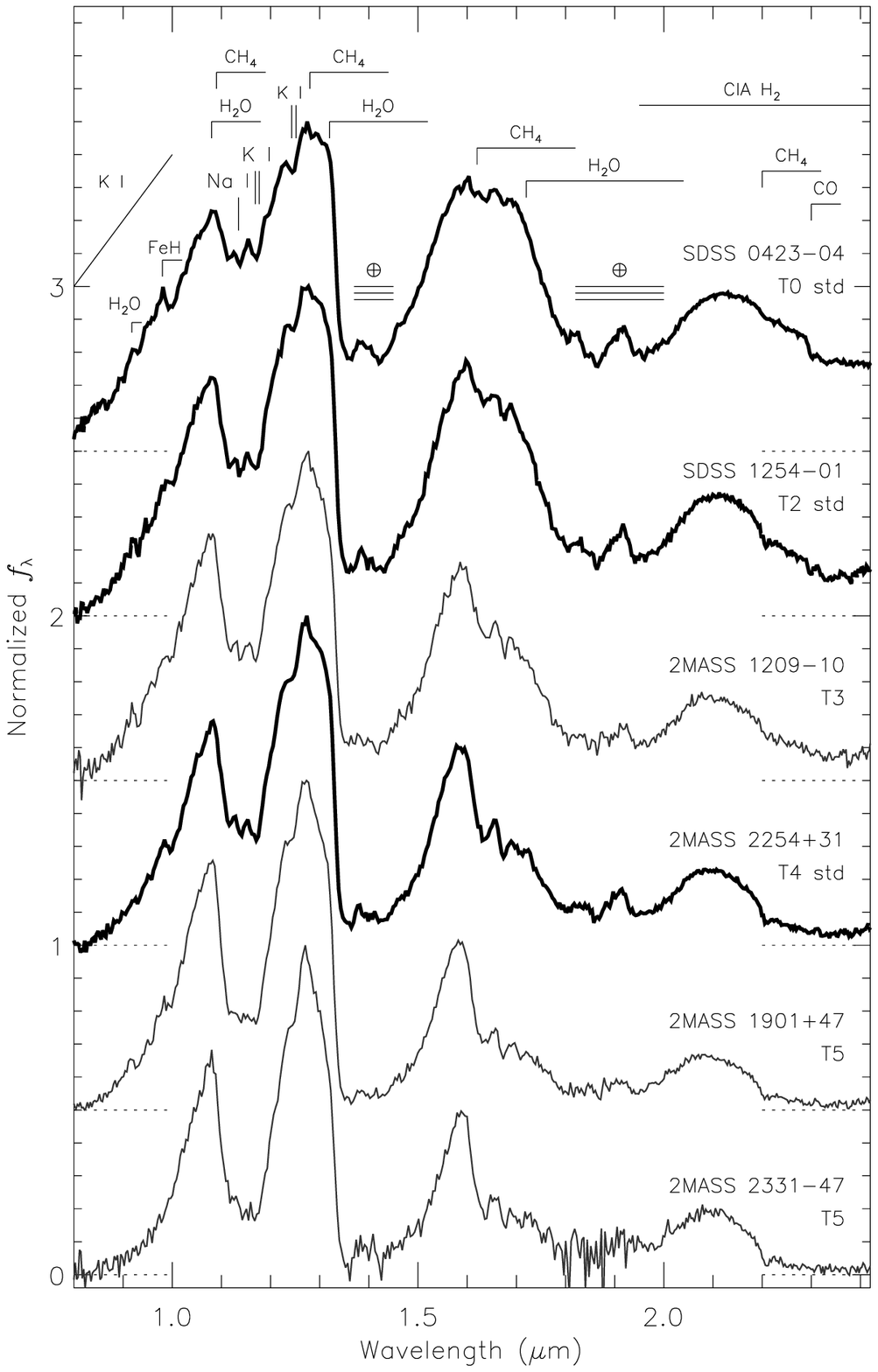}{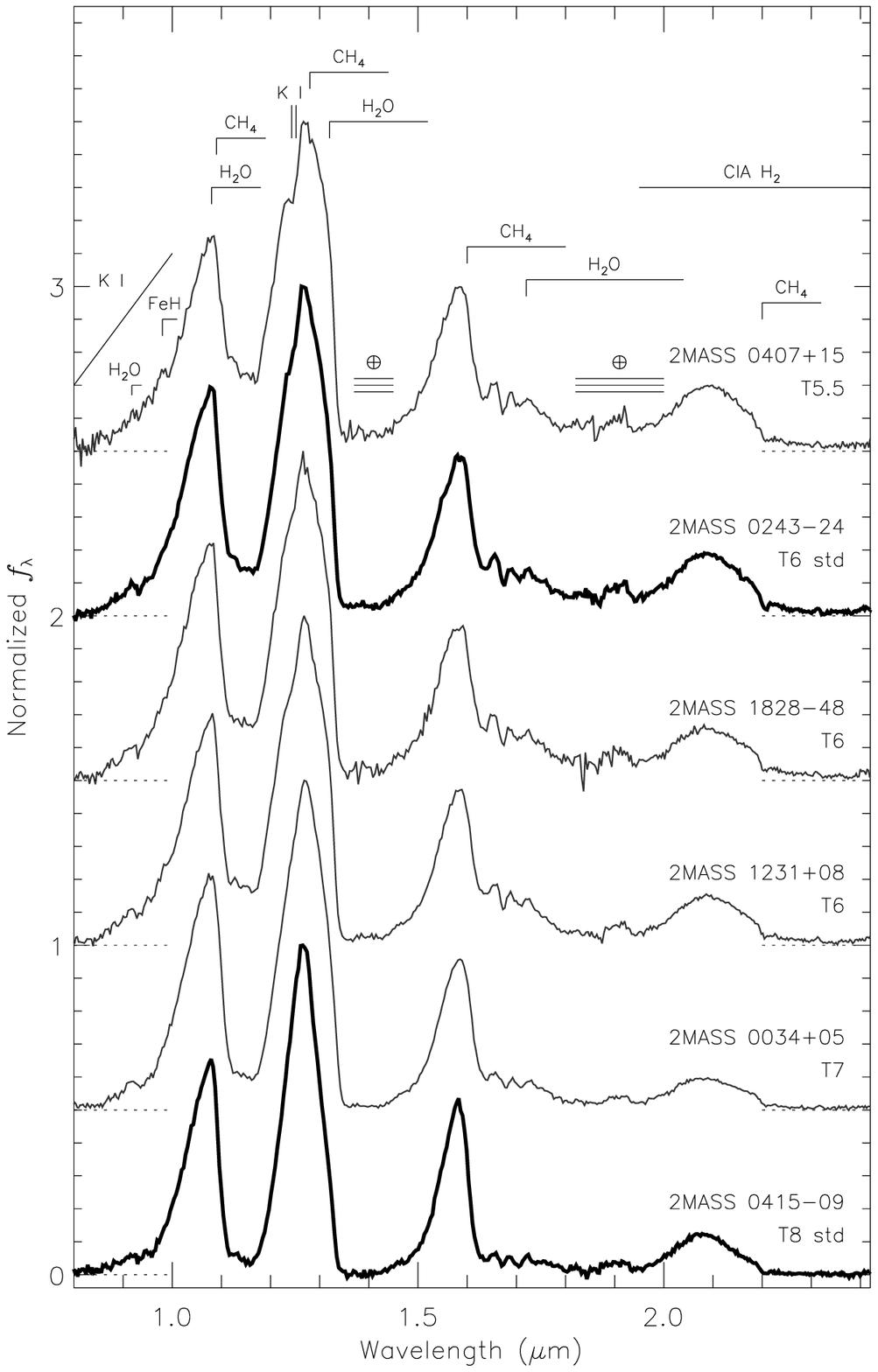}
\caption{SpeX prism spectra of T dwarf discoveries (thin grey lines) and spectral standards
(thick black lines).  Spectra
are normalized at 1.28 $\micron$ and offset by a constant (dotted lines).  Key spectral features
are indicated, as are regions of strong telluric
absorption ($\earth$).}
\end{figure}

\begin{figure}
\epsscale{0.3}
\centering
\plotone{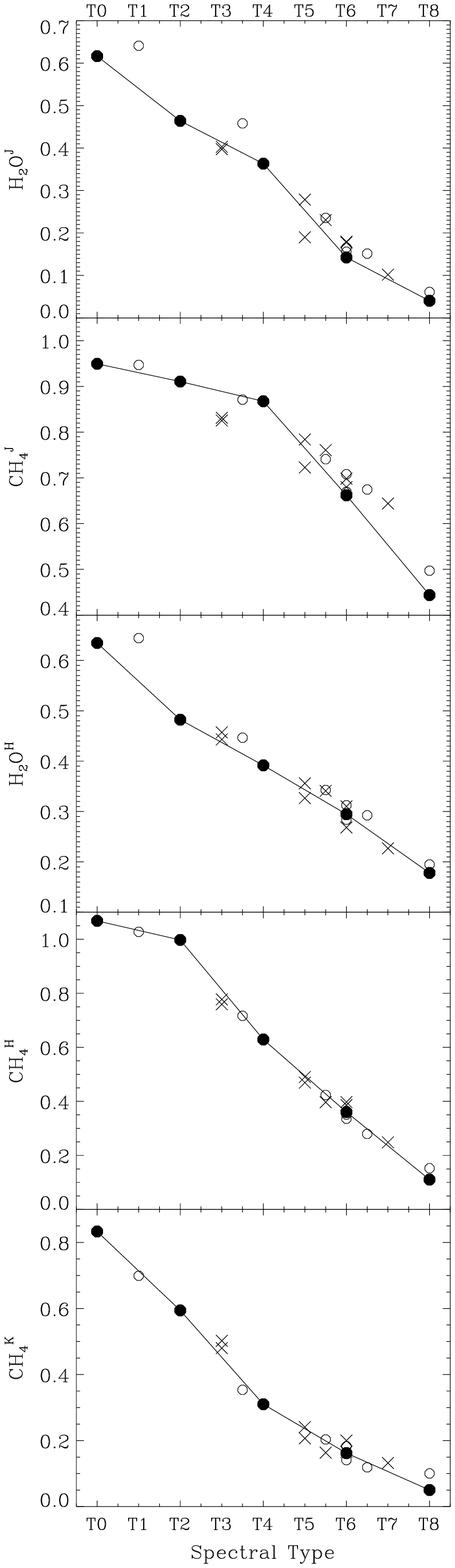}
\caption{Spectral indices versus spectral type.  Solid circles connected by
lines indicate spectral standard values, open circles indicate previously known
and classified T dwarfs, and crosses indicate T dwarf discoveries from this paper.
Spectral types are those in Table 6, with assigned spectral types for the standards,
previously published spectral types for the known T dwarfs, and derived spectral types
for the new discoveries.}
\end{figure}

\begin{figure}
\epsscale{1.0}
\centering
\plotone{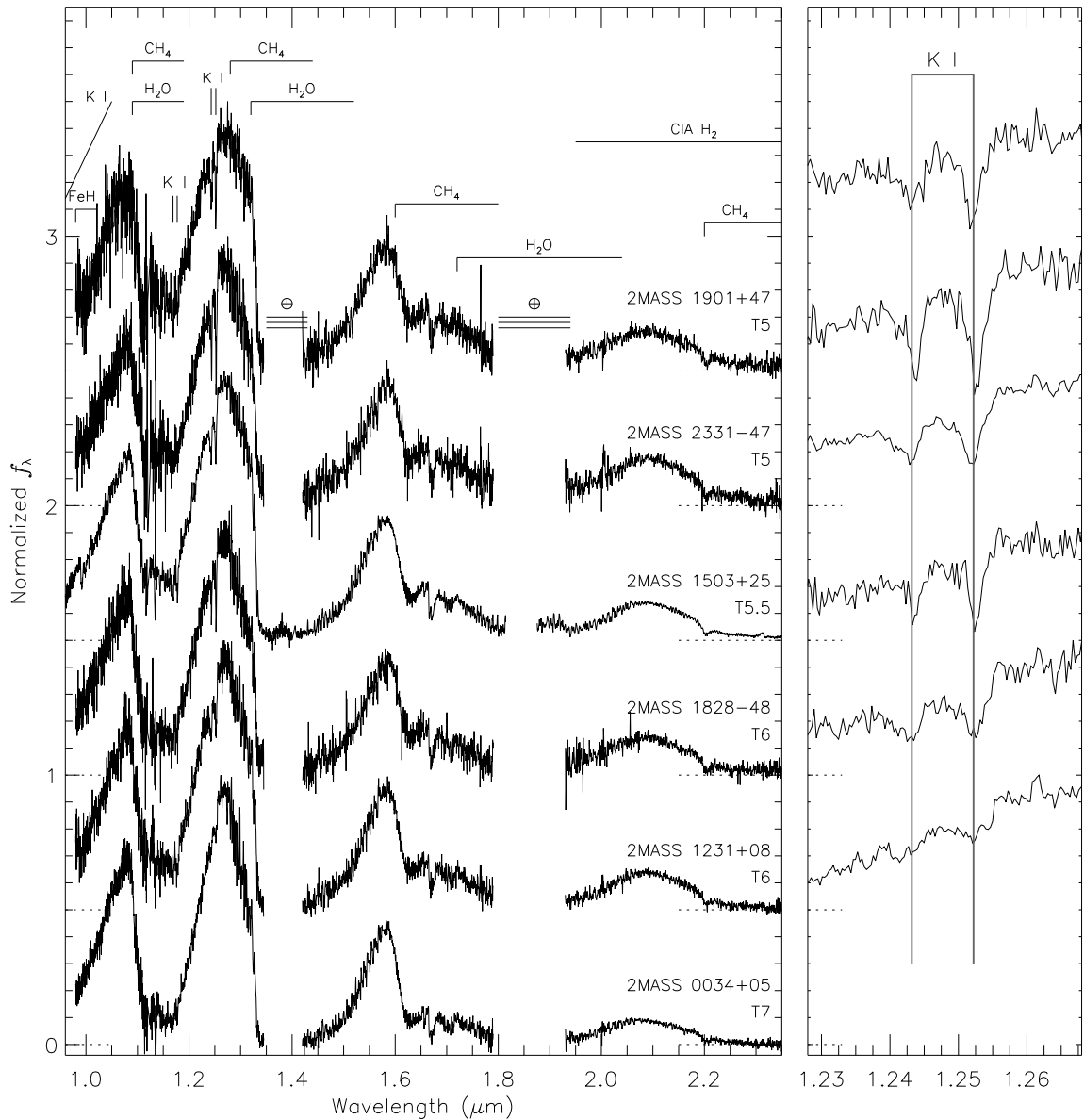}
\caption{Moderate-resolution (R$\sim$1200) SpeX spectra for six T dwarfs observed.
({\em Left}) Full spectra,
normalized at 1.28 $\micron$ and offset by a constant (dotted lines).
Regions of strong telluric absorption at 1.4 and 1.9 $\micron$ are omitted with
the exception of the bright source 2MASS 1503+2525, where
the gap between 1.79 and 1.84 $\micron$ is due to order separation.
Features are noted as in Figure 7. ({\em Right}) Close-up of the spectral
region around the 1.243/1.252 $\micron$ \ion{K}{1} doublet.  Spectra are scaled and offset
as in the left panel.}
\end{figure}

\begin{figure}
\epsscale{1.0}
\centering
\plotone{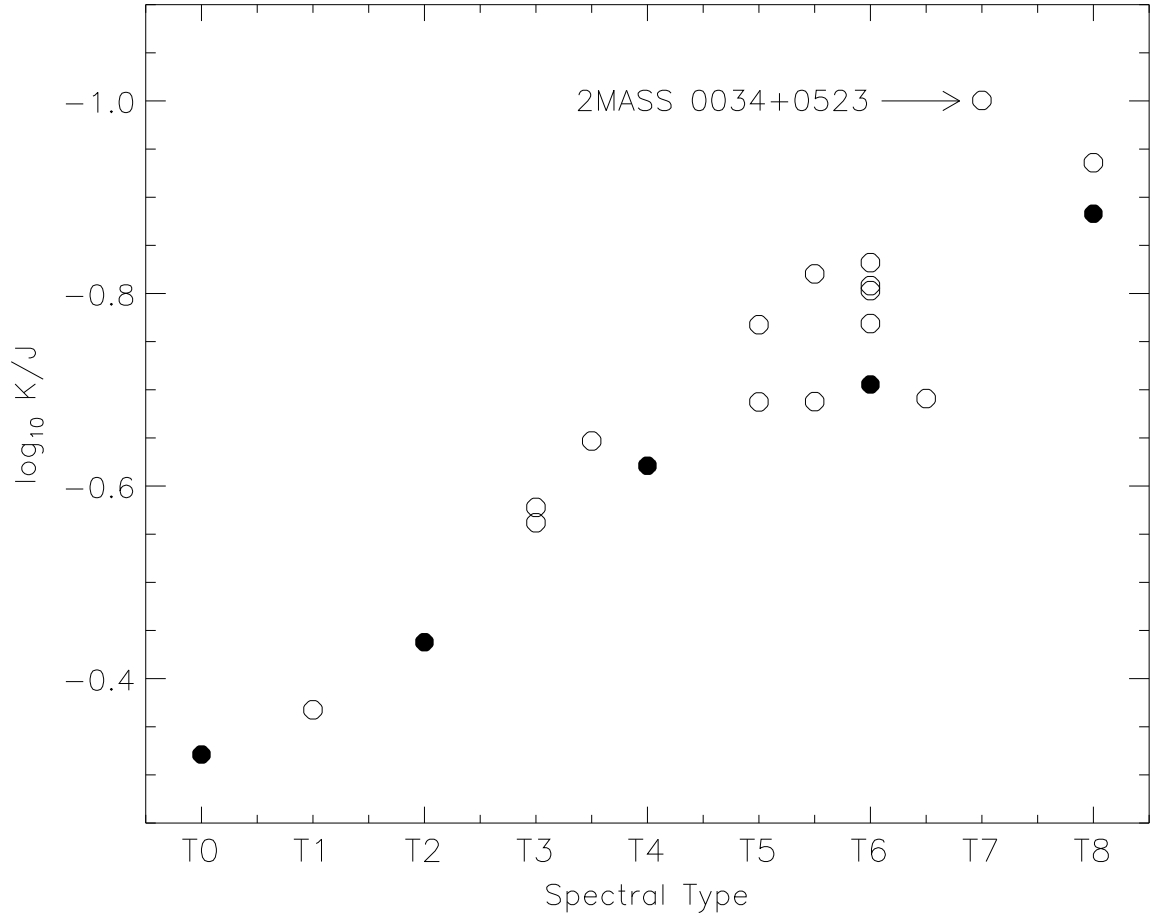}
\caption{Spectral index $\log_{10}{K/J}$ versus T spectral type for objects observed with the SpeX
low-resolution prism mode.  Spectral standards are indicated by solid circles, all others are indicated
by open circles.  Note the excellent linear correlation across the full spectral type range,
with the exception of 2MASS 0034+0523 (indicated) which exhibits a more suppressed $K$-band peak due
to enhanced CIA H$_2$ absorption.}
\end{figure}

\end{document}